 \documentclass[final,1p,times,onecolumn,authoryear]{elsarticle}

\usepackage{amssymb}
\usepackage{lipsum}
\usepackage{subcaption}
\usepackage{graphicx}
\usepackage{float}
\usepackage{hyperref}

\journal{ }

\begin{document}

\begin{frontmatter}

\title{The role of Solar Activity in shaping Precipitation Extremes: A Regional Exploration in Kerala, India}

\author[inst1]{Elizabeth Thomas}
\ead{shinuelz@yahoo.co.in}

\affiliation[inst1]{organization={Department of Physics},
           addressline={Mar Thoma College}, 
           city={ Kuttapuzha P. O., Tiruvalla },
            postcode={ PIN 689103}, 
            state={Kerala},
            country={India}}
            
\author[inst1]{S. Vineeth}
\ead{vineethsmaikkattu2@gmail.com}

\author[inst1]{Noble P. Abraham\corref{cor1}}
 \ead{noblepa@gmail.com}

 \cortext[cor1]{Corresponding author}
 
\begin{abstract}
There has been global attention focused on extreme climatic changes. The purpose of this paper is to explore the response of extreme precipitation events to solar activity, over Kerala, India. The three solar indices—sunspot number, F10.7 index, and cosmic ray intensity—are examined, and their relationship to rainfall is examined during a 57-year period (1965–2021), starting with Solar Cycle 20. Both solar and rainfall data are considered on an annual scale as well as on a seasonal scale by dividing them into winter, pre-monsoon, monsoon, and post-monsoon seasons. The solar indices are used to calculate correlation coefficients with seasonal rainfall. Through correlation analysis, it is found that the precipitation in Kerala is correlated with the sunspot activity, but with different significance. When solar activity is high, the winter and monsoon seasons exhibit strong correlations with high significance. The solar influence at the regional level is also studied. The central and southern parts of Kerala appear to be influenced by the Sun during periods of high activity. The years with excess and deficiency of rainfall are calculated and compared with the solar indices. It was observed that the years with excessive and insufficient rainfall coincide with the years when the solar activity is at its highest or minimum. It is suggested that there is a physical link and a way to predict extreme rainfall events in Kerala based on the association between solar activity and those events.
\end{abstract}


\end{frontmatter}




\section{Introduction}
\label{introduction}

Global climate change poses a hazard to human existence. A significant influence on weather and climate is exerted by the sun and anthropogenic factors. The Sun's magnetic fields exhibit a wide range of spatial, temporal, and energetic phenomena. Sunspots, solar flares, solar wind, coronal mass ejections, etc. are all expressions of magnetic activity in the Sun, collectively known as solar activity  \citep{Usoskin2017}. Sunspot number, solar radio flux, and cosmic rays are some of the widely used solar indices used to quantify solar activity. Sunspot number quantifies sunspots and is widely used because of its long-term availability. Sunspot number is highly correlated with other solar indices \citep{hathaway2015solar,tiwari2018}. A measure of solar radio flux at 10.7 cm is called the F10.7 index, which originates deep in the corona and high in the chromosphere \citep{tapping1994limits,tapping201310}. Cosmic rays are high-energy particles that travel to the Earth from outside the solar system. Cosmic rays have been found to have a negative correlation with the number of sunspots \citep{Gupta2006}.

There has long been concern about how the sun affects precipitation on Earth. Precipitation in different parts of the world appears to be affected by the sun at different time intervals. The effect of solar activity on precipitation varies with time scale and region, leading to different correlations \citep{Tsiropoula2003,ZhaoJuanHanYan-BenandLi2004,Wasko2009,Mauas2011,Rampelotto2012}. Recently, few studies have been conducted on the relationship between solar and precipitation in China \citep{Zhai2017,YU2019,yan2022}, the United States \citep{Nitka2019}, Europe \citep{Laurenz2019}, Argentina \citep{HEREDIA2019105094}, Nepal \citep{tiwari2021}, and Northeast Asia \citep{yan2022}.

The economy, agriculture, and ecosystem in India could be seriously impacted by changing rainfall patterns \citep{DoranaluChandrashekar2017}. Many researchers have looked into the potential of a connection between solar activity and rainfall throughout India or in various regions \citep{jagannathan1973changes,Ananthakrishnan1984,Hiremath2004,Bhattacharyya2005,Agnihotri2011,Badruddin2015,Warrier2017,Thomas2022}. The direct and indirect effects were studied and the results were often localised and contradict other authors \citep{JAGANNATHAN1973,bhalme1981solar,Hiremath2006,Bhattacharyya2007,Lihua2007,Selvaraj2009,selvaraj2011study,Selvaraj2013,Hiremath2015,Malik2018,Thomas2023}.

\begin{figure*}[!t]
\centering
\includegraphics[width=0.75\textwidth]{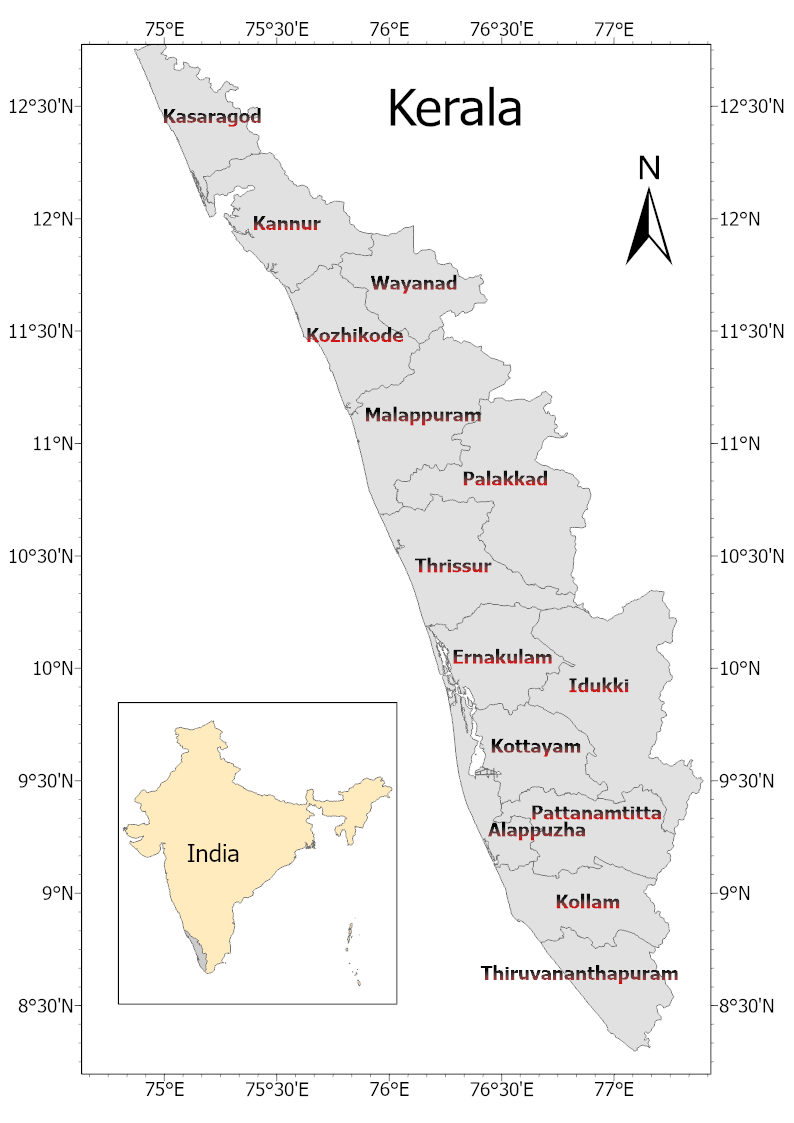}
\caption{Location map of Kerala}
\label{kerala}
\end{figure*}

Kerala is located at the southwest tip of India, bounded on the east by the Western Ghats and the west by the Arabian Sea. It extends between 8$^{\circ}$15$^{\prime}$ and 12$^{\circ}$50$^{\prime}$ north latitudes
and between 74$^{\circ}$50$^{\prime}$ and 77$^{\circ}$30$^{\prime}$ east longitudes. It shares boundaries with Karnataka in the
north, Tamil Nadu in the east, and the Arabian Sea in the west. Kerala
has a wet and tropical climate and the major contribution is from the
southwest monsoon and post-monsoon. The diverse features of Kerala make it more susceptible to climate change. It is known as the "Gateway of summer monsoon". Studies of long-term rainfall variability revealed that rainfall during the southwest monsoon significantly reduced while rainfall during the post-monsoon rose \citep{Krishnakumar2009,Kothawale2017}. Recently, few studies have reported the influence of sunspot number with the rainfall over Kerala \citep{thomas2022impact,Thomas2022,Thomas2023}. The location map of Kerala is given in Figure \ref{kerala}.

In Kerala, recent extreme rainfall events have resulted in landslides or floods that have claimed lives and destroyed property. In India, several studies have linked solar activity to extreme weather events (see for e.g. \cite{bhalme1981cyclic,Azad2011}). Therefore, it will be interesting to look at extreme rainfall over the Kerala region with various solar parameters.

Observing rainfall's response to different solar parameters might reveal subtle differences. In this paper, the influence of different solar parameters, ie, sunspot number, F10.7 Index and cosmic ray intensity, on extreme precipitation over Kerala is evaluated using correlative studies.
Section \ref{data} discusses the data and methodology of analysis. Section \ref{results} presents the results and discussion about the correlation and variation of different solar parameters with rainfall over Kerala. It also includes results regarding the occurrences of extreme rainfall events. Section \ref{conclusions} presents the conclusions.

\begin{figure*}[!t]
\centering
\includegraphics[width=0.75\textwidth]{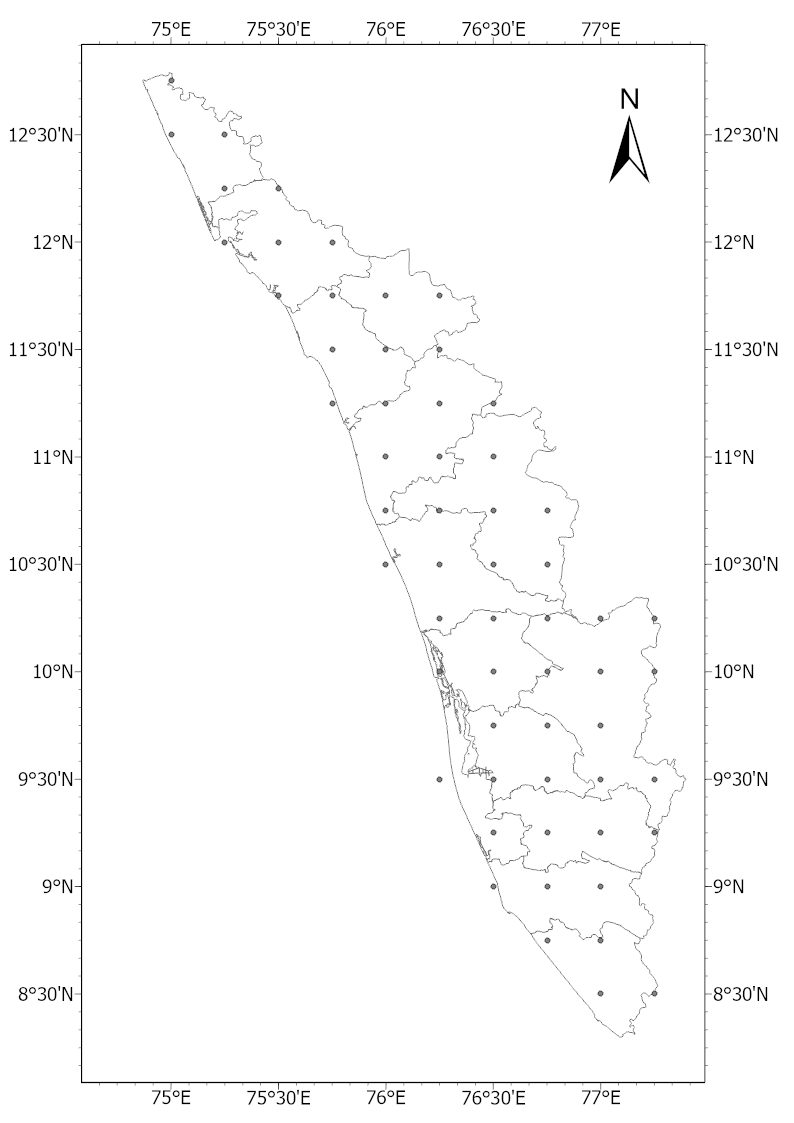}
\caption{Location of Grid points}
\label{keralagrid}
\end{figure*}

\section{Data and methodology of analysis}
\subsection{Datasets}
\label{data}
We have used 57 years (1965–2021) of data on rainfall in Kerala as well as solar indices (sunspot number, F10,7 Index, and cosmic ray intensity). This covers the time starting with Solar Cycle 20. The sunspot number data was taken from the World Data Center SILSO, Royal Observatory of Belgium, Brussels. The F10.7 Index, solar flux data (in sfu, 1 sfu = $10^{-22} W m^{-2} Hz^{-1}$) was obtained from LASP Interactive Solar Irradiance Data Center.
 The Oulu Cosmic Ray station provided data on cosmic ray intensity (measured in counts/min). Rainfall (in mm) over Kerala was obtained from the India Meteorological Department’s
(IMD) daily gridded rainfall dataset of high spatial resolution
(0.25$^{\circ}$ × 0.25$^{\circ}$) \citep{Pai2014}. In this study, 59 grid locations from across the districts in Kerala were taken into consideration, as shown in Figure \ref{keralagrid}.

\begin{figure*}[!t]
\centering
  \begin{subfigure}{0.50\textwidth}
    \includegraphics[width=\linewidth]{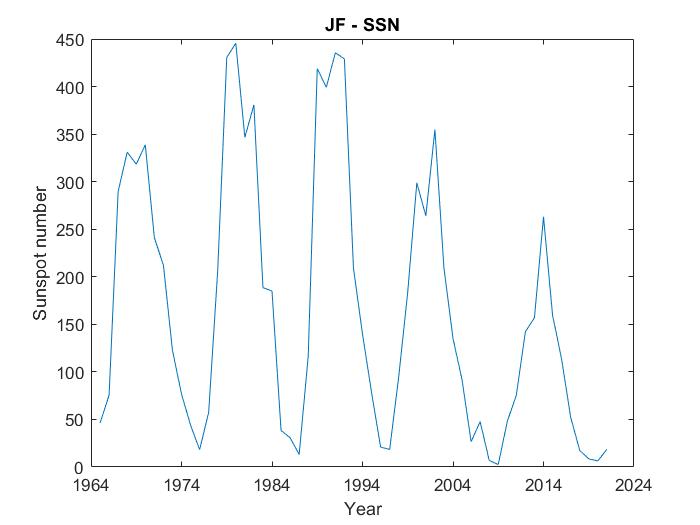}
    \caption{} \label{fig:3a}
  \end{subfigure}%
  \begin{subfigure}{0.50\textwidth}
    \includegraphics[width=\linewidth]{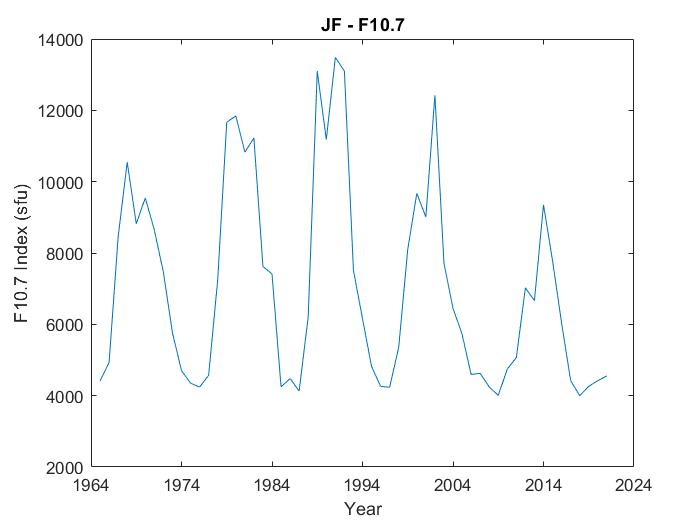}
    \caption{} \label{fig:3b}
  \end{subfigure}  \\
   \begin{subfigure}{0.50\textwidth}
    \includegraphics[width=\linewidth]{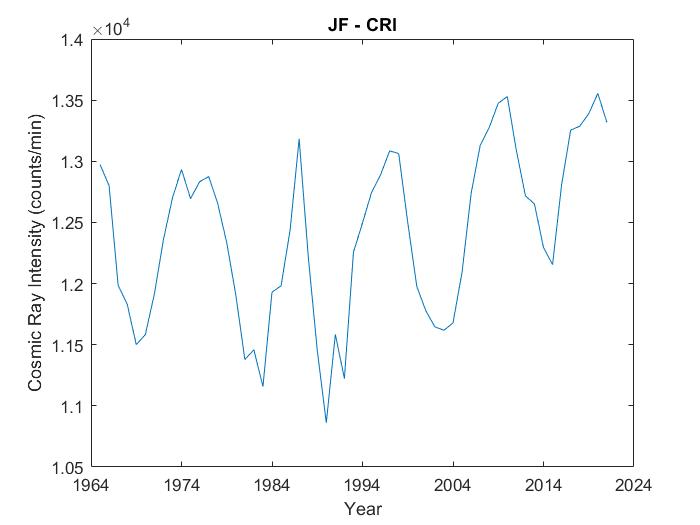}
    \caption{} \label{fig:3c}
  \end{subfigure}%
  \begin{subfigure}{0.50\textwidth}
    \includegraphics[width=\linewidth]{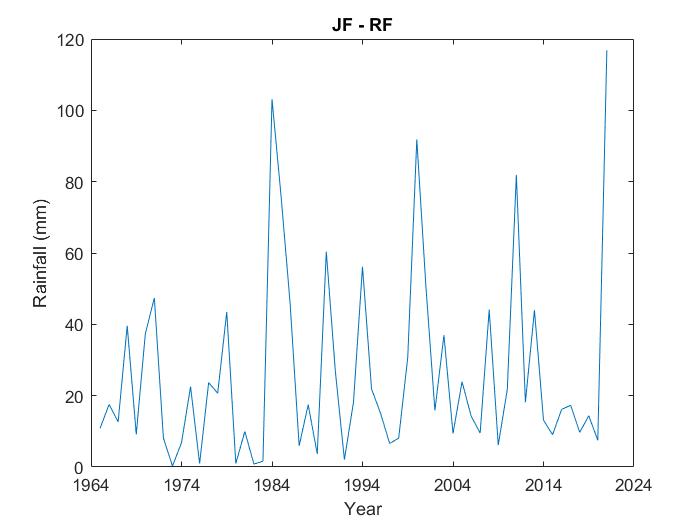}
    \caption{} \label{fig:3d}
  \end{subfigure}  \\
\caption{Time series of (a) sunspot number (SSN) (b) F10.7 Index (F10.7) (c) Cosmic ray intensity (CRI) and (d) rainfall (RF), corresponding to JF season.}
\label{jfseason}
\end{figure*}

\begin{figure*}[!t]
\centering
  \begin{subfigure}{0.50\textwidth}
    \includegraphics[width=\linewidth]{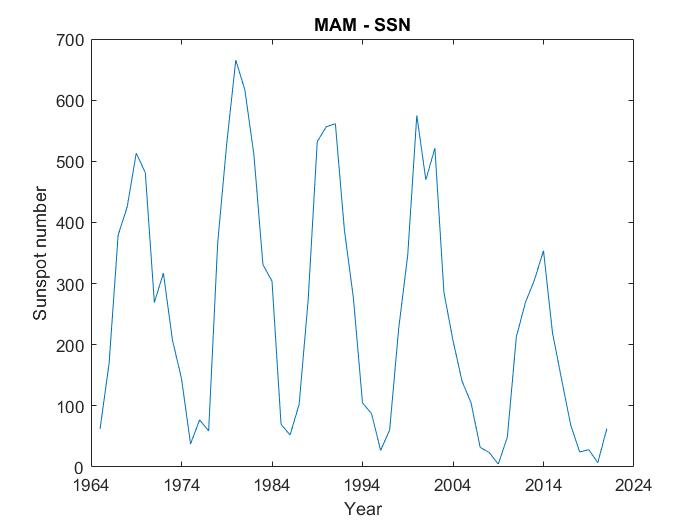}
    \caption{} \label{fig:4a}
  \end{subfigure}%
  \begin{subfigure}{0.50\textwidth}
    \includegraphics[width=\linewidth]{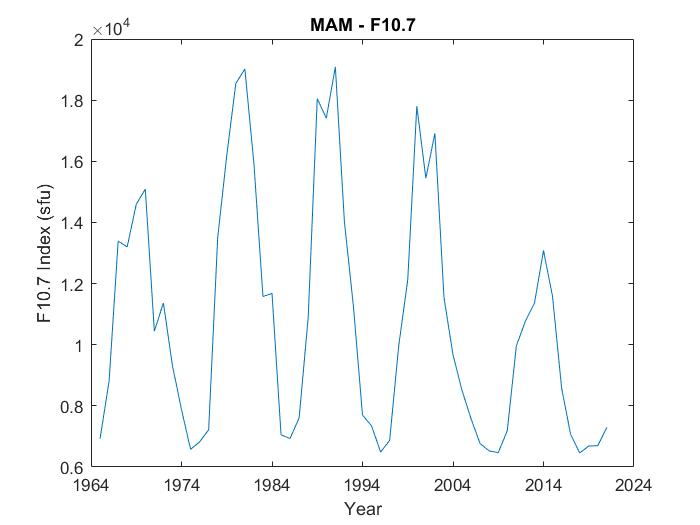}
    \caption{} \label{fig:4b}
  \end{subfigure}  \\
   \begin{subfigure}{0.50\textwidth}
    \includegraphics[width=\linewidth]{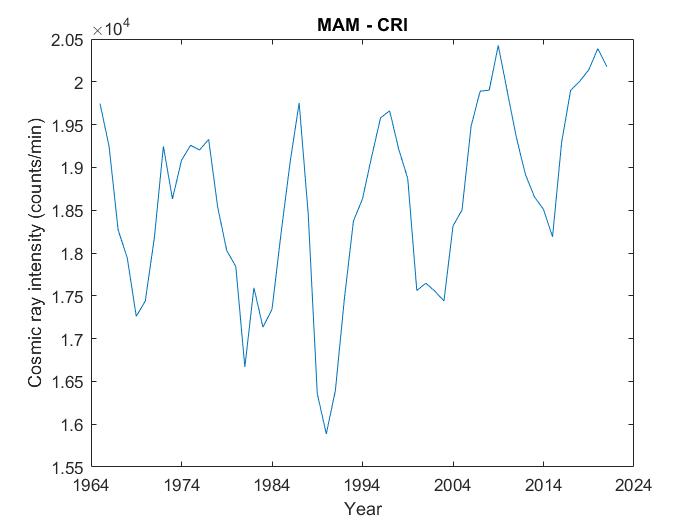}
    \caption{} \label{fig:4c}
  \end{subfigure}%
  \begin{subfigure}{0.50\textwidth}
    \includegraphics[width=\linewidth]{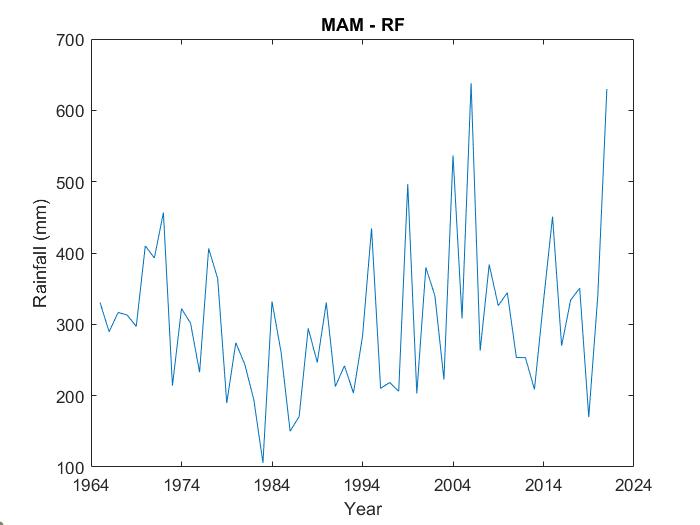}
    \caption{} \label{fig:4d}
  \end{subfigure}  \\
\caption{Time series of (a) sunspot number (SSN) (b) F10.7 Index (F10.7) (c) Cosmic ray intensity (CRI) and (d) rainfall (RF), corresponding to MAM season.}
\label{mamseason}
\end{figure*}

\begin{figure*}[!t]
\centering
  \begin{subfigure}{0.50\textwidth}
    \includegraphics[width=\linewidth]{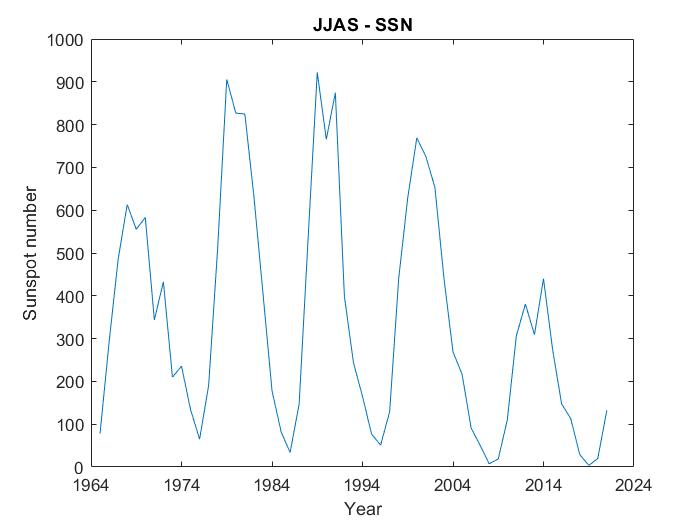}
    \caption{} \label{fig:5a}
  \end{subfigure}%
  \begin{subfigure}{0.50\textwidth}
    \includegraphics[width=\linewidth]{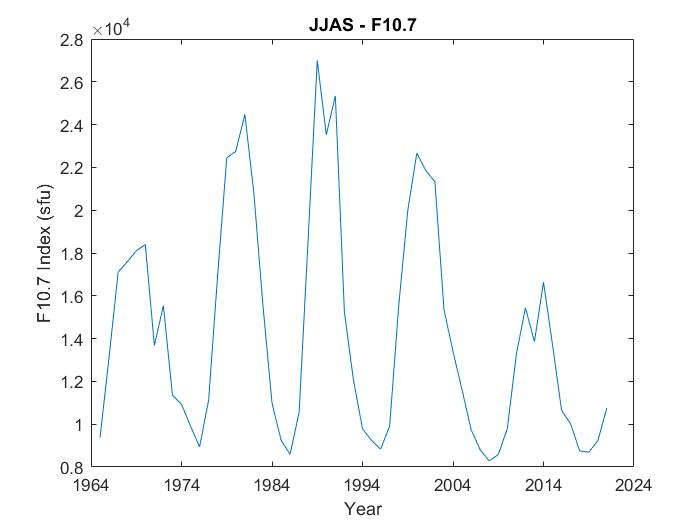}
    \caption{} \label{fig:5b}
  \end{subfigure}  \\
   \begin{subfigure}{0.50\textwidth}
    \includegraphics[width=\linewidth]{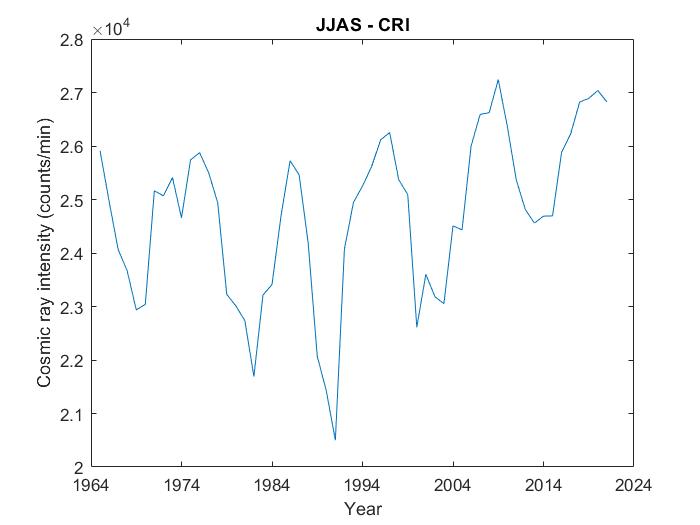}
    \caption{} \label{fig:5c}
  \end{subfigure}%
  \begin{subfigure}{0.50\textwidth}
    \includegraphics[width=\linewidth]{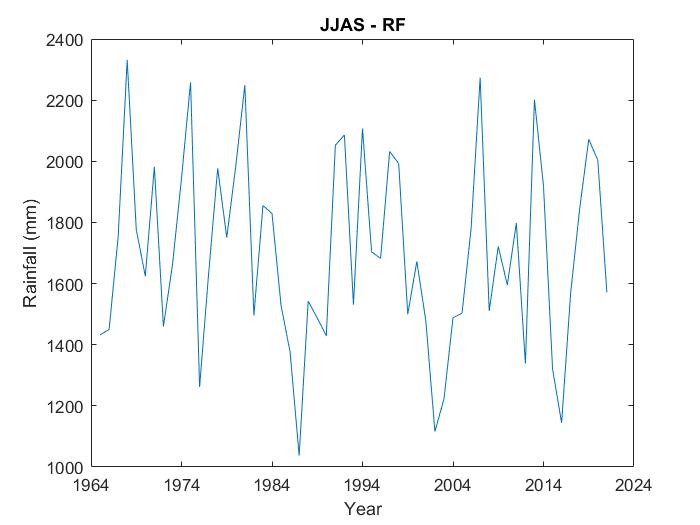}
    \caption{} \label{fig:5d}
  \end{subfigure}  \\
\caption{Time series of (a) sunspot number (SSN) (b) F10.7 Index (F10.7) (c) Cosmic ray intensity (CRI) and (d) rainfall (RF), corresponding to JJAS season.}
\label{jjasseason}
\end{figure*}

\begin{figure*}[!t]
\centering
  \begin{subfigure}{0.50\textwidth}
    \includegraphics[width=\linewidth]{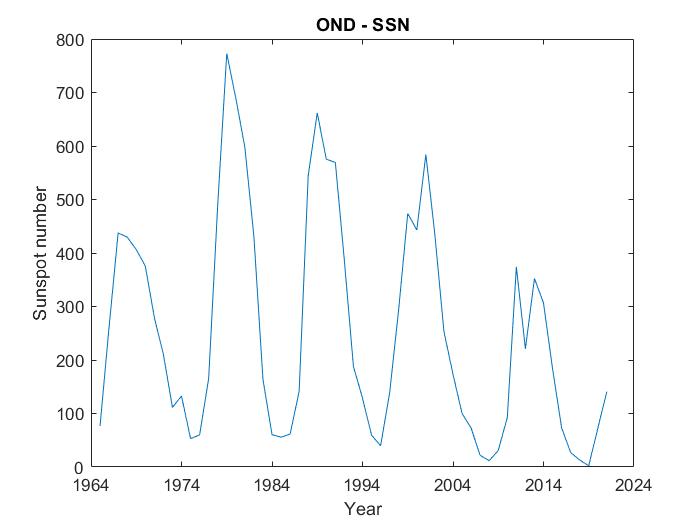}
    \caption{} \label{fig:6a}
  \end{subfigure}%
  \begin{subfigure}{0.50\textwidth}
    \includegraphics[width=\linewidth]{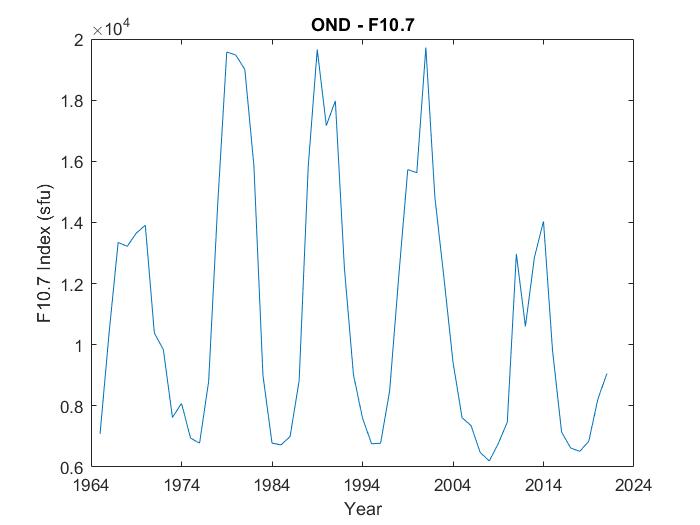}
    \caption{} \label{fig:6b}
  \end{subfigure}  \\
   \begin{subfigure}{0.50\textwidth}
    \includegraphics[width=\linewidth]{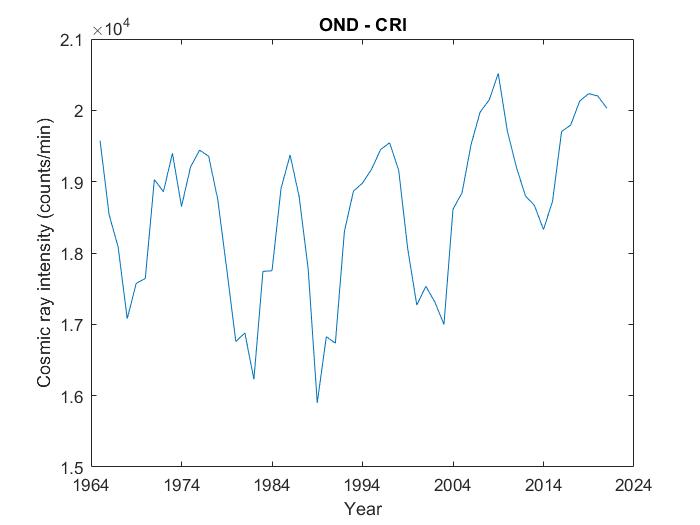}
    \caption{} \label{fig:6c}
  \end{subfigure}%
  \begin{subfigure}{0.50\textwidth}
    \includegraphics[width=\linewidth]{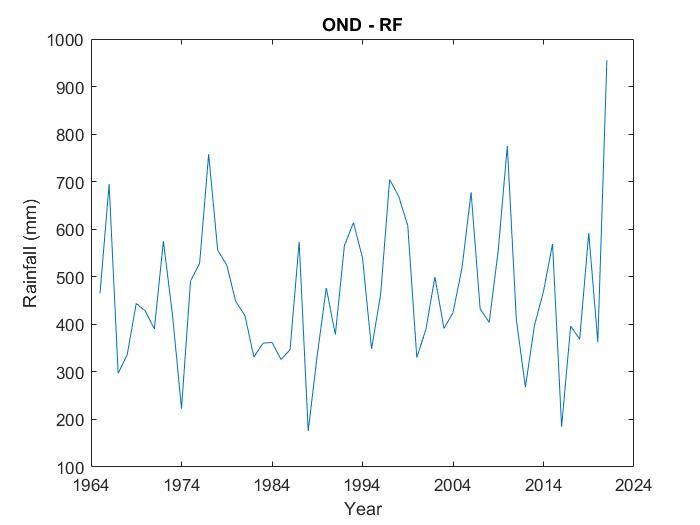}
    \caption{} \label{fig:6d}
  \end{subfigure}  \\
\caption{Time series of (a) sunspot number (SSN) (b) F10.7 Index (F10.7) (c) Cosmic ray intensity (CRI) and (d) rainfall (RF), corresponding to OND season.}
\label{ondseason}
\end{figure*}

India Meteorological Department (IMD) classifies the rainfall seasons of India as Winter (January-February), pre-monsoon (March-May), monsoon (June-September), and post-monsoon (October-December) \citep{Hiremath2004, Hiremath2006, Bankoti2011}. In this study, the rainfall and the different solar indices were grouped into these four seasons, and the corresponding values for sunspot number are given as SSN, F10.7 Index is given as F10.7 (sfu), cosmic ray intensity is given as CRI (count/min), and rainfall is given as RF (mm). 
 Figure \ref{jfseason}, Figure \ref{mamseason}, Figure \ref{jjasseason} and Figure \ref{ondseason} show the time series of SSN, F10.7, CRI, and RF corresponding to the JF, MAM, JJAS and OND seasons respectively.
 
\subsection{Methodology of analysis} \label{methodology}
To understand the response of precipitation over Kerala with the solar indices (SSN, F10.7, and CRI), correlation coefficients were calculated. Here, the period of high solar activity, ie, Solar Cycle 21 (1977-1986) was taken into consideration and the rainfall and solar indices data were grouped accordingly. Correlation coefficients are usually computed to check whether any relationship exists between the two data sets and how strong the relationship is. Here, Spearman Rank-Order correlation coefficients and their significance were calculated to determine the relationship between different solar activity indices (SSN, F10.7, and CRI) and rainfall RF data. This method of correlation is more powerful than linear correlation \citep{Hiremath2004,Hiremath2006,Bankoti2011}. The correlation coefficients corresponding to the 59 grid points were determined and the results were represented as correlation maps.

The years of excess and deficit rainfall over Kerala were identified to study the relationship between extreme rainfall events and solar activity. For that, the mean ($\mu$) and standard deviation ($\sigma$) of rainfall RF, during both the seasons (JF and JJAS), were determined. A year i was labelled as extreme rainfall year when $R_{i} \geq (\mu + \sigma)$ and a year labelled as deficient rainfall year when 
$R_{i} \leq (\mu + \sigma)$, where $R_{i}$ is the rainfall of that year, $ i,k\in\mathbb{R} $ \citep{Azad2011}. In this study, k was defined as one. 

\section{Results and Discussions}\label{results}
\subsection{Relationship between Sunspot number and Rainfall}

\begin{figure*}[!t]
\centering
  \begin{subfigure}{0.5\textwidth}
    \includegraphics[width=\linewidth]{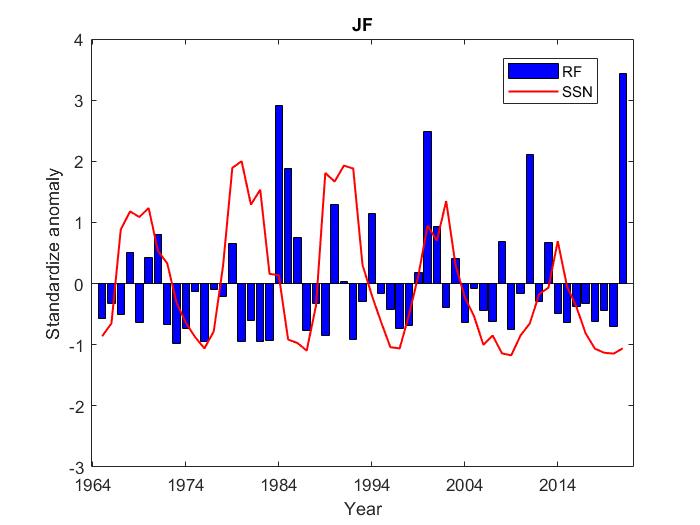}
    \caption{} \label{fig:7a}
  \end{subfigure}%
  \begin{subfigure}{0.5\textwidth}
    \includegraphics[width=\linewidth]{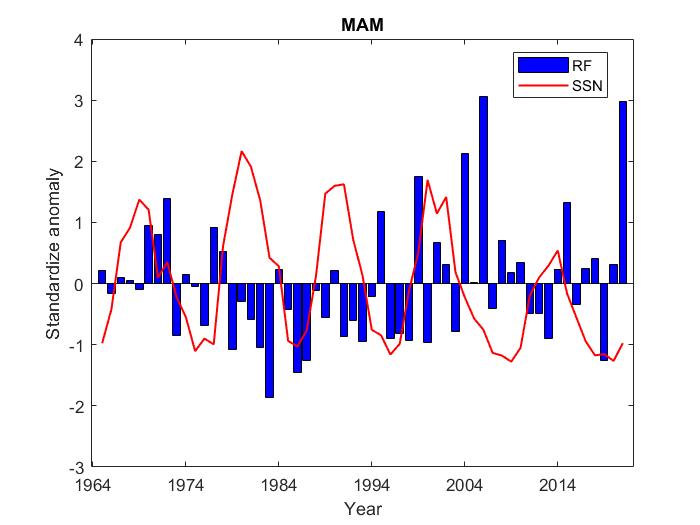}
    \caption{} \label{fig:7b}
  \end{subfigure}  \\
  \begin{subfigure}{0.5\textwidth}
    \includegraphics[width=\linewidth]{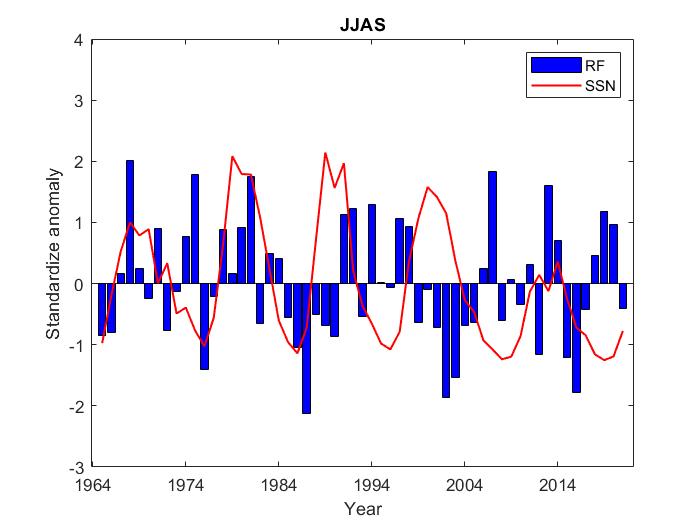}
    \caption{} \label{fig:7c}
  \end{subfigure}%
  \begin{subfigure}{0.5\textwidth}
    \includegraphics[width=\linewidth]{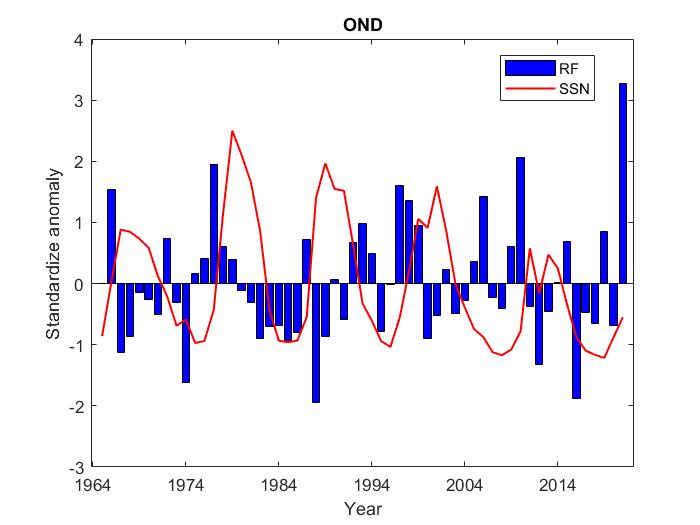}
    \caption{} \label{fig:7d}
  \end{subfigure}  \\
\caption{Variation of sunspot number (SSN) and rainfall (RF) during (a) JF (b) MAM (c) JJAS and (d) OND seasons.}
\label{ssnrf}
\end{figure*} 

\begin{figure*}[p]
\centering
  \begin{subfigure}{0.45\textwidth}
    \includegraphics[width=\linewidth]{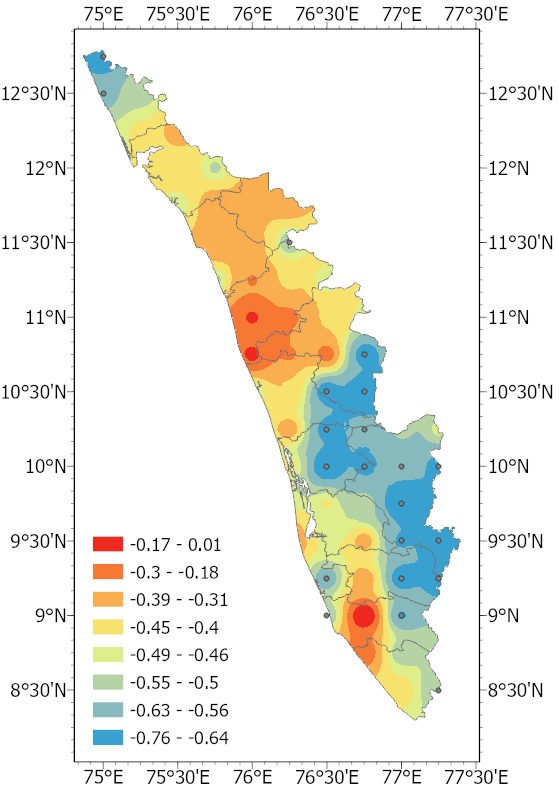}
    \caption{JF} \label{fig:8a}
  \end{subfigure}%
  \begin{subfigure}{0.45\textwidth}
    \includegraphics[width=\linewidth]{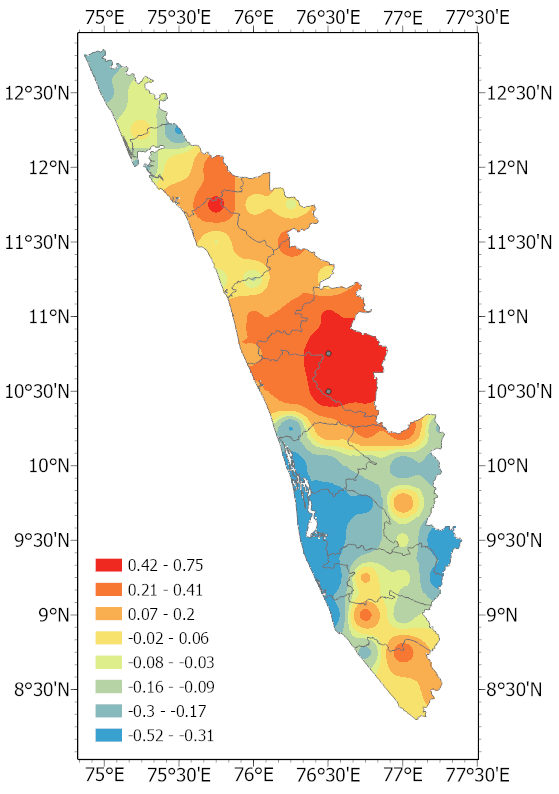}
    \caption{MAM} \label{fig:8b}
  \end{subfigure}  \\
  \begin{subfigure}{0.45\textwidth}
    \includegraphics[width=\linewidth]{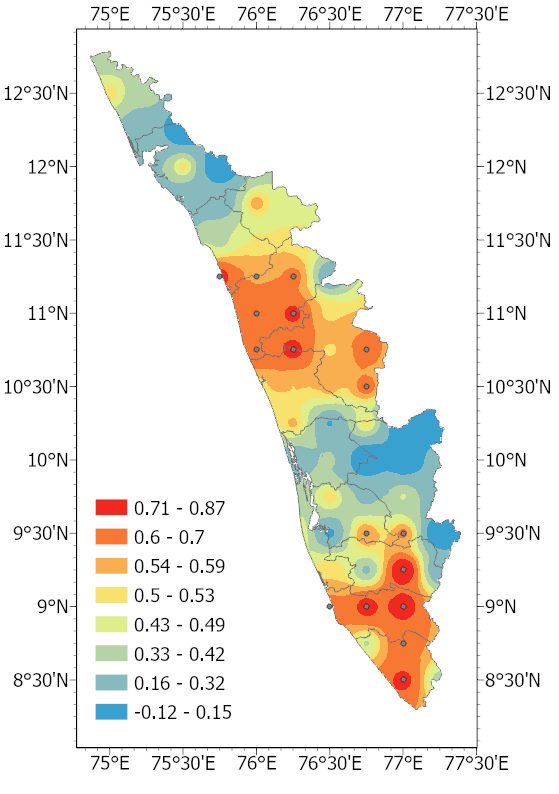}
    \caption{JJAS} \label{fig:8c}
  \end{subfigure}%
  \begin{subfigure}{0.45\textwidth}
    \includegraphics[width=\linewidth]{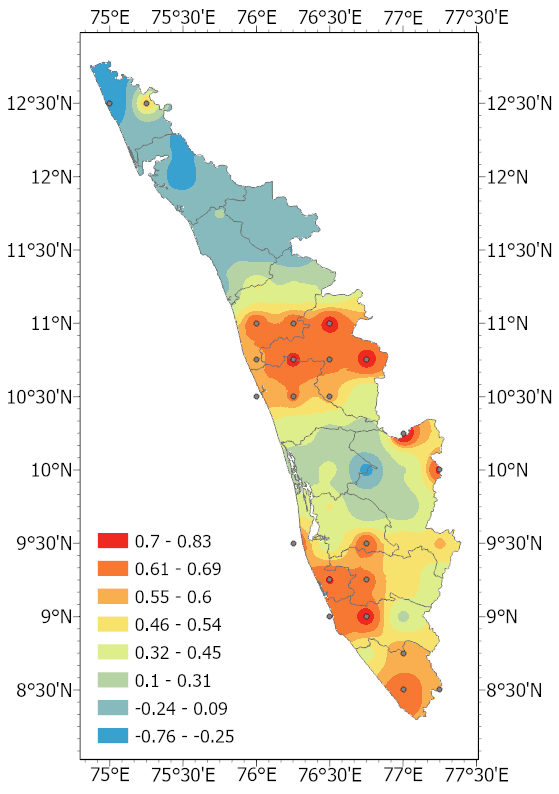}
    \caption{OND} \label{fig:8d}
  \end{subfigure}  \\
\caption{Correlation coefficients of Sunspot number (SSN) and rainfall over Kerala (RF) during (a) JF (b) MAM (c) JJAS and (d) OND seasons. (Dark dots represents grid points with correlations at 0.1 significance level.) }
        \label{corr_ssn}
\end{figure*}

Figure \ref{ssnrf} shows the time series of standardized values of sunspot number (SSN) and rainfall (RF) during the JF, MAM, JJAS, and OND seasons.  To study the response of rainfall over Kerala to the sunspot number during high solar activity, the Spearman rank-order correlation coefficients between SSN and RF corresponding to each grid point were calculated and the results displayed as a correlation map, given in Figure \ref{corr_ssn}. During the JF season, rainfall is negatively correlated with the sunspot number throughout the region. A highly negative correlation, with significance, is observed at the topmost region, ie, in parts of Kasargod and the eastern boundary regions covering Palakkad, Idukki, Pathanamthitta, Kollam, and Ernakulam districts. The MAM rainfall shows a positive correlation, with significance, with sunspot number in the central part of Kerala which becomes strong in areas mainly covering Palakkad district. In the case of the JJAS season, the central and the southern regions of Kerala showed a positive and significant correlation, while the remaining regions gave weaker results. The OND season revealed weaker correlation results compared to other seasons. A significant positive correlation was visible in the central and the south-western regions covering Palakkad, Thrissur, Malappuram, Alappuzha, Kollam, and Thiruvananthapuram districts. The northern districts till Kozhikode showed a relative negative correlation during the JJAS and OND seasons.

The precipitation over Kerala was noted to be correlated with sunspot number, with varying significance, irrespective of signs. Variations in correlation results were observed during all the seasons. The JF and JJAS seasons revealed a stronger relationship between rainfall and sunspot number, compared to other seasons.

\subsection{Relationship between F10.7 Index and Rainfall}

\begin{figure*}[t!]
\centering
  \begin{subfigure}{0.5\textwidth}
    \includegraphics[width=\linewidth]{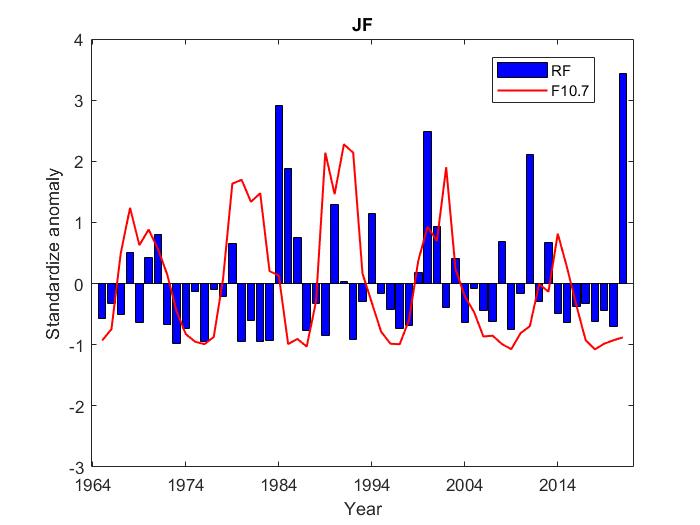}
    \caption{} \label{fig:9a}
  \end{subfigure}%
  \begin{subfigure}{0.5\textwidth}
    \includegraphics[width=\linewidth]{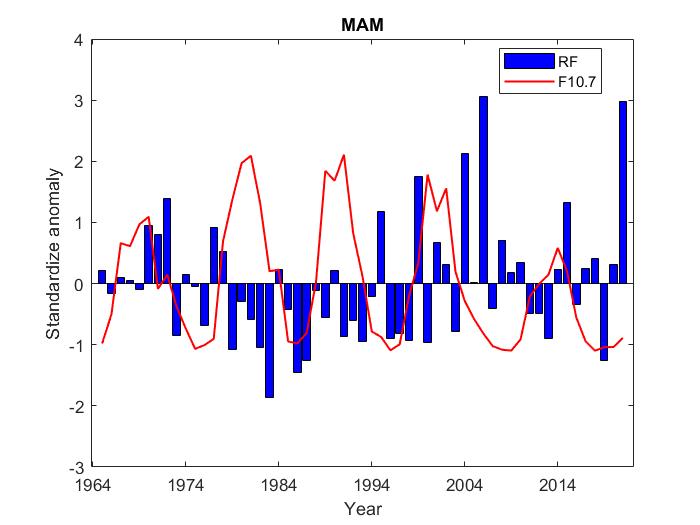}
    \caption{} \label{fig:9b}
  \end{subfigure}  \\
  \begin{subfigure}{0.5\textwidth}
    \includegraphics[width=\linewidth]{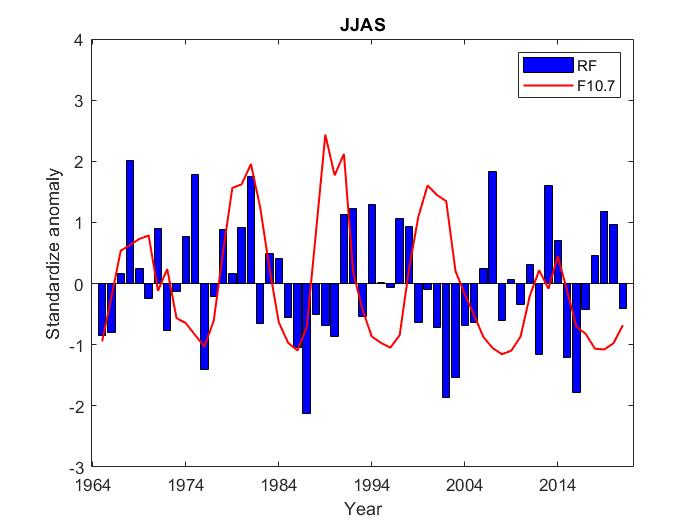}
    \caption{} \label{fig:9c}
  \end{subfigure}%
  \begin{subfigure}{0.5\textwidth}
    \includegraphics[width=\linewidth]{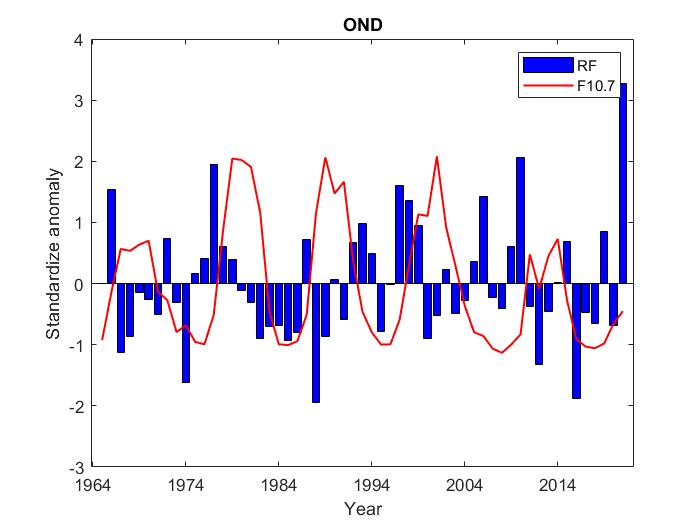}
    \caption{} \label{fig:9d}
  \end{subfigure}  \\
\caption{Variation of F10.7 Index (F10.7) and rainfall (RF) during (a) JF (b) MAM (c) JJAS and (d) OND seasons.}
\label{f107rf}
\end{figure*} 

\begin{figure*}[p]
\centering
  \begin{subfigure}{0.45\textwidth}
    \includegraphics[width=\linewidth]{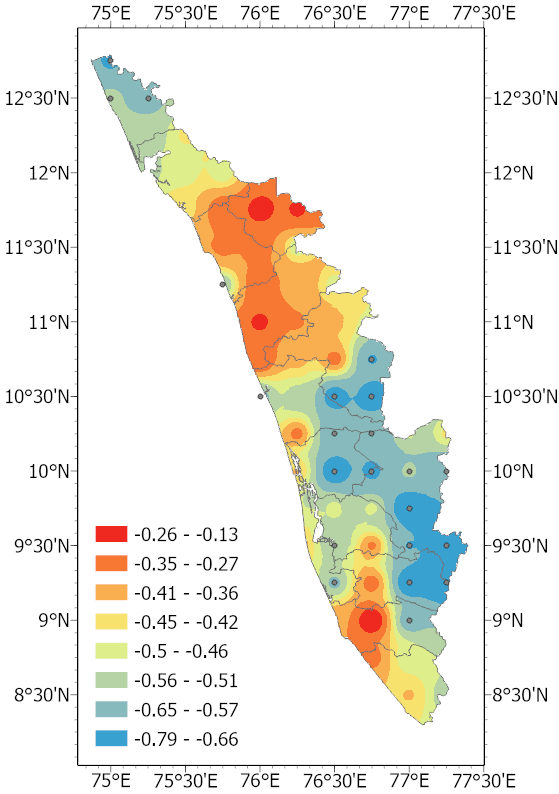}
    \caption{JF} \label{fig:10a}
  \end{subfigure}%
  \begin{subfigure}{0.45\textwidth}
    \includegraphics[width=\linewidth]{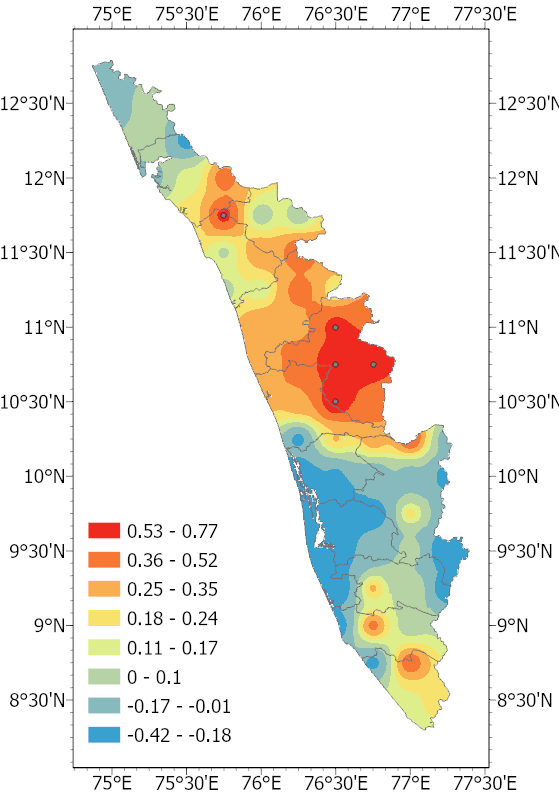}
    \caption{MAM} \label{fig:10b}
  \end{subfigure}  \\
  \begin{subfigure}{0.45\textwidth}
    \includegraphics[width=\linewidth]{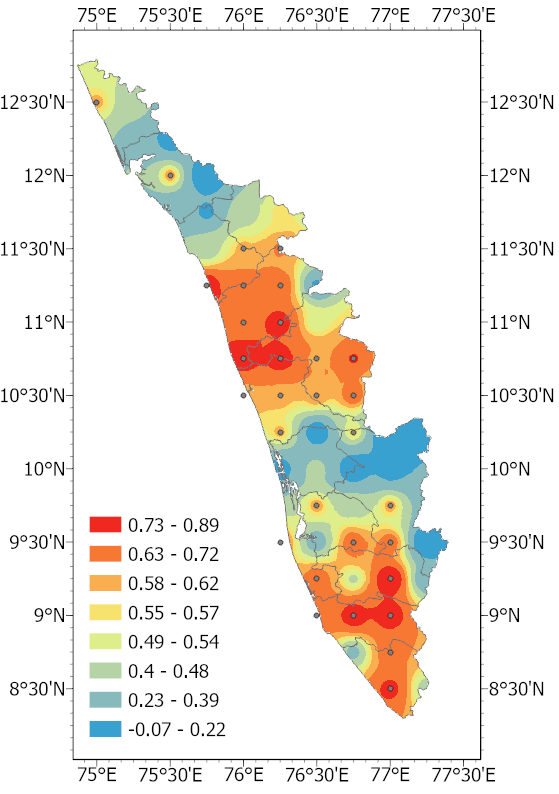}
    \caption{JJAS} \label{fig:10c}
  \end{subfigure}%
  \begin{subfigure}{0.45\textwidth}
    \includegraphics[width=\linewidth]{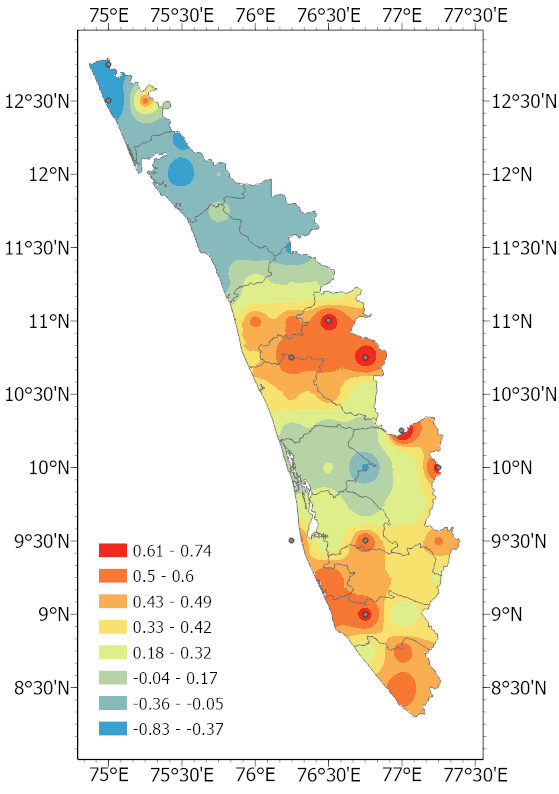}
    \caption{OND} \label{fig:10d}
  \end{subfigure}  \\
\caption{Correlation coefficients of F10.7 Index (F10.7) and rainfall over Kerala (RF) during (a) JF (b) MAM (c) JJAS and (d) OND seasons. (Dark dots represents grid points with correlations at 0.1 significance levels.) }
        \label{corr_f107}
\end{figure*}

A time series of standardized values of F10.7 Index (F10.7) and rainfall (RF) for the JF, MAM, JJAS, and OND seasons is represented by Figure \ref{f107rf}.
The Spearman rank-order correlation coefficients between F10.7 and RF were calculated to assess the dependence of rainfall over Kerala on F10.7 Index during high solar activity. The results are given in Figure \ref{corr_f107}. Correlation results were similar to SSN results, though weaker. Similar to the SSN results, the JF season showed a negative correlation between F10.7 Index and rainfall over the entire region. The northernmost region and the eastern regions extending towards the south showed a significant negative correlation. The MAM rainfall displayed a significant positive correlation with the F10.7 Index in the central parts of the state, which is similar to that of the SSN. In the case of the JJAS season, F10.7 Index showed better correlation results than that of SSN. A high positive correlation with significance was observed mainly in the central regions of Kerala, covering mainly Malappuram, Palakkad, Thrissur districts and Pathanamthitta, Alappuzha, Kollam, and Thiruvananthapuram districts in the south. Compared to the other seasons, the OND season revealed weaker correlation results.

\subsection{Relationship between Cosmic ray intensity and Rainfall}

\begin{figure*}[t!]
\centering
  \begin{subfigure}{0.5\textwidth}
    \includegraphics[width=\linewidth]{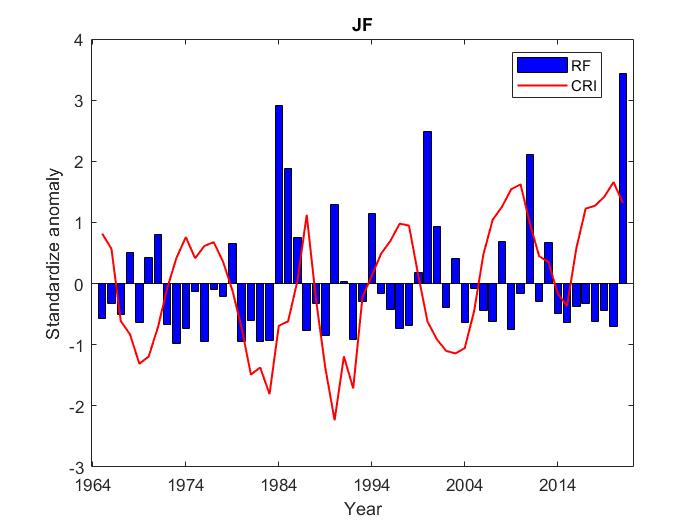}
    \caption{} \label{fig:11a}
  \end{subfigure}%
  \begin{subfigure}{0.5\textwidth}
    \includegraphics[width=\linewidth]{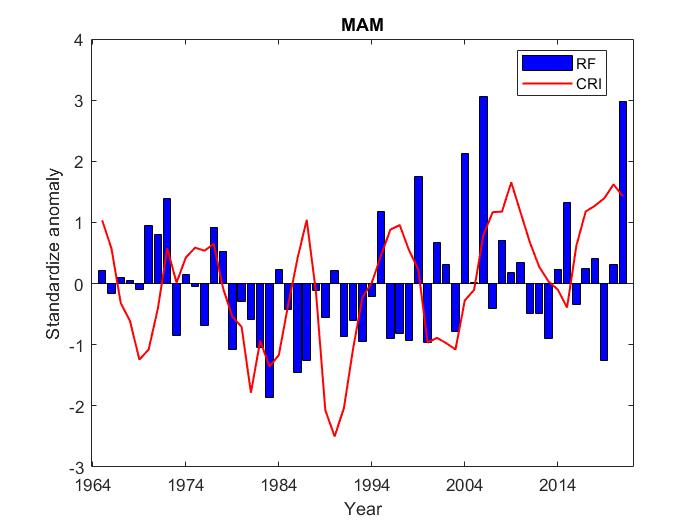}
    \caption{} \label{fig:11b}
  \end{subfigure}  \\
  \begin{subfigure}{0.5\textwidth}
    \includegraphics[width=\linewidth]{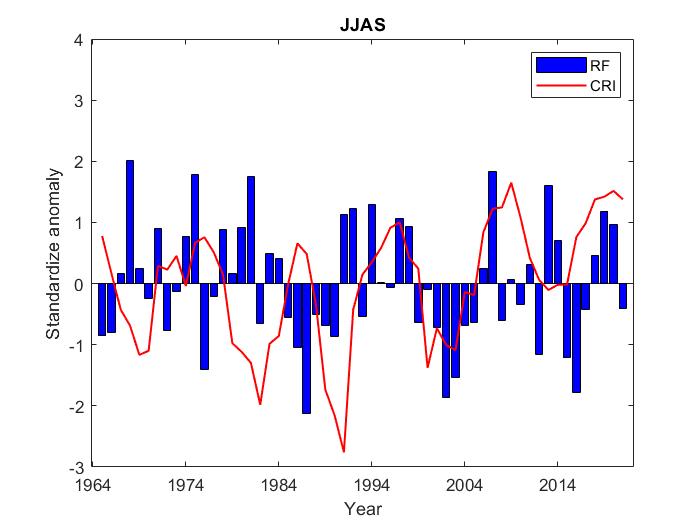}
    \caption{} \label{fig:11c}
  \end{subfigure}%
  \begin{subfigure}{0.5\textwidth}
    \includegraphics[width=\linewidth]{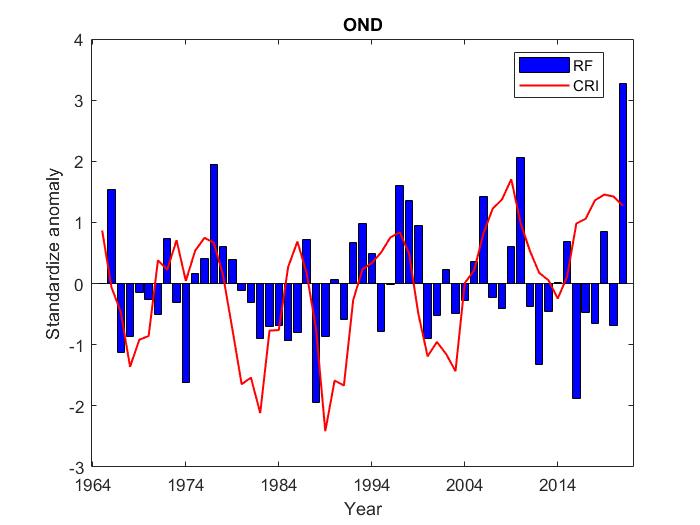}
    \caption{} \label{fig:11d}
  \end{subfigure}  \\
\caption{Variation of Cosmic Ray Intensity (CRI) and rainfall (RF) during (a) JF (b) MAM (c) JJAS and (d) OND seasons.}
\label{crirf}
\end{figure*} 

Figure \ref{crirf} shows the time series of standardized values of Cosmic ray intensity (CRI) and rainfall (RF) during different seasons. 

\begin{figure*}[p]
\centering
  \begin{subfigure}{0.45\textwidth}
    \includegraphics[width=\linewidth]{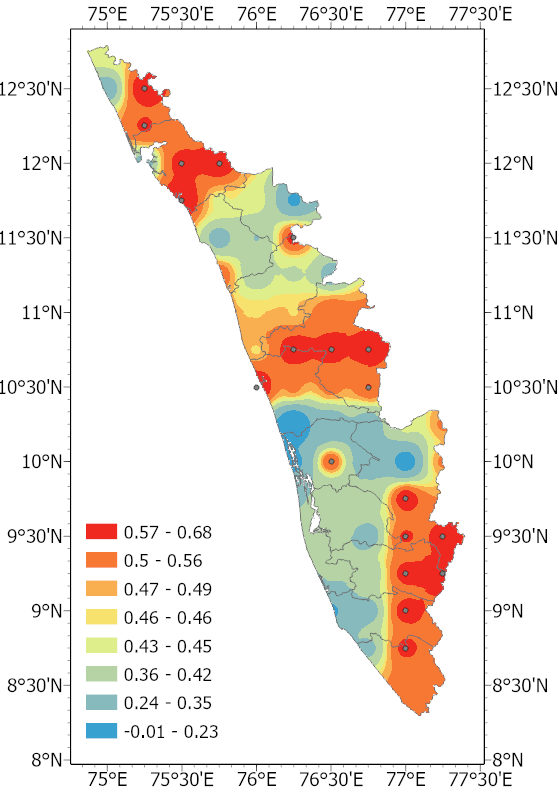}
    \caption{JF} \label{fig:12a}
  \end{subfigure}%
  \begin{subfigure}{0.45\textwidth}
    \includegraphics[width=\linewidth]{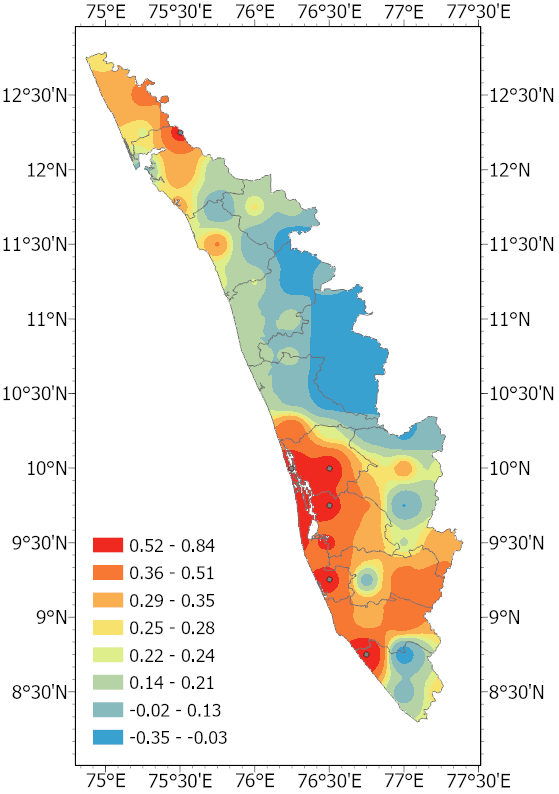}
    \caption{MAM} \label{fig:12b}
  \end{subfigure}  \\
  \begin{subfigure}{0.45\textwidth}
    \includegraphics[width=\linewidth]{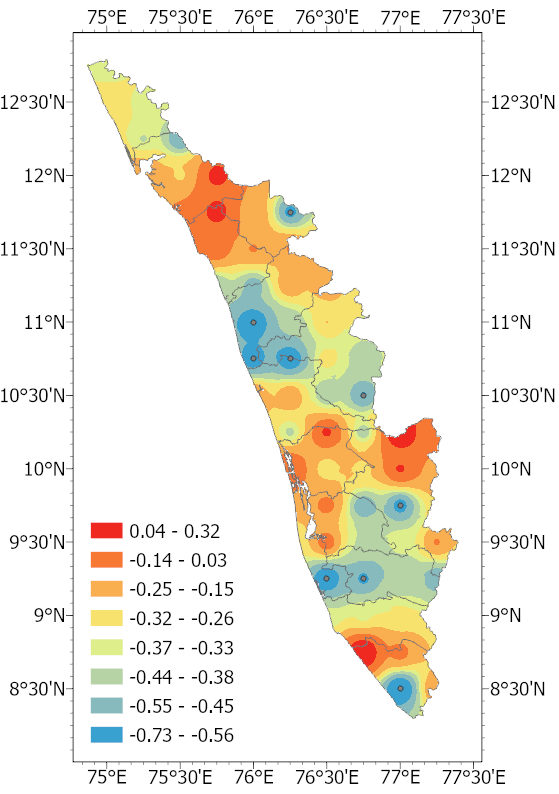}
    \caption{JJAS} \label{fig:12c}
  \end{subfigure}%
  \begin{subfigure}{0.45\textwidth}
    \includegraphics[width=\linewidth]{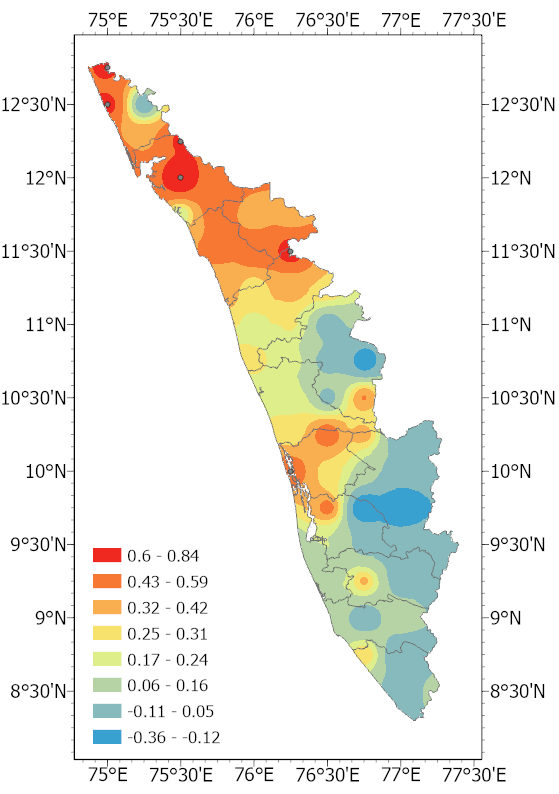}
    \caption{OND} \label{fig:12d}
  \end{subfigure}  \\
\caption{Correlation coefficients of Cosmic ray intensity (CRI) and rainfall over Kerala (RF) during (a) JF (b) MAM (c) JJAS and (d) OND seasons. (Dark dots represents grid points with correlations at 0.1 significance levels.) }
        \label{corr_cri}
\end{figure*}

Spearman rank-order correlation coefficients between CRI and RF were calculated to examine the influence of solar activity on rainfall over Kerala. The results are given in Figure \ref{corr_cri}. The correlation results observed were completely different from those of SSN and F10.7 results. The influence of cosmic ray intensity on the rainfall over Kerala is weaker compared to other solar indices. During the JF season, positive correlation with significance was observed in different parts of the state, mainly covering Kasargod, Kannur, Palakkad, Thrissur, and along the eastern boundary, ie, parts of Idukki, Pathanamthitta, Kollam, and Thiruvananthapuram districts. The MAM rainfall showed a positive correlation with significance along the southwestern regions covering Ernakulam, Kottayam, Alappuzha, and Thiruvananthapuram districts. The JJAS results could not draw many significant conclusions. In the case of OND season, the northern regions showed a positive correlation while coming to the south, the correlation sign reverses and becomes negative. 

The rainfall over Kerala was observed to be influenced by the solar activity with varying significance, during different seasons. Out of the three solar indices considered, the sunspot number and F10.7 Index showed better correlation compared to cosmic ray intensity. The JF and JJAS seasons were observed to be more related to solar activity than the other two seasons. The central and southern parts of Kerala seem to be affected by the sun during high-activity periods. The rainfall received in different parts of the state varies during different seasons. Kerala's geographical conditions, terrain, and rainfall variations may be contributing factors to the varying correlation results.

Solar influences on precipitation have been investigated in several correlative studies in India. According to \cite{Ananthakrishnan1984}, an analysis of 306 Indian stations revealed both positive and negative correlation coefficients.
 \cite{Hiremath2004} investigated the correlation between sunspot number on seasonal and annual Indian monsoon rainfall. He found that pre-monsoon and monsoon rainfall showed significant positive correlations. \cite{Hiremath2006} examined correlative effects of sunspot number over Indian rainfall during each solar cycle and found a moderate to high significance correlation irrespective of the sign. When solar activity was low, rainfall was higher than when solar activity was high. A weak positive and negative correlation during different seasons was reported by \cite{Bal2010}. In the study of solar parameters (Sunspot number, solar active prominence, and H alpha solar flares) and All India homogeneous rainfall, \cite{Bankoti2011}  found that the correlations varied in their sign according to the season and with the solar parameters. A seasonal correlation between cosmic rays and rainfall was observed by \cite{Chaudhuri2015} during the post-monsoon season. Many correlative studies were performed in different states to find a possible sun-rainfall link, ie, in West Bengal \citep{chakraborty1986solar}, Rajasthan \citep{jain1997correlation}, Tamil Nadu \citep{selvaraj2009influence,selvaraj2011study,selvaraj2012} and Kerala \citep{thomas2022impact}.
 
 Similar correlative studies were recently performed globally, attempting to link rainfall with solar activity in China \citep{yan2022}, the United States \citep{Nitka2019}, and Europe \citep{Laurenz2019}.

\subsection{Solar indices and extreme rainfall in Kerala}

Attempts were made to understand the possible relationship between solar activity and extreme precipitation events in Kerala during different seasons (JF, MAM, JJAS, and OND). For that, the years of excess rainfall and deficient rainfall were identified, as explained in Section \ref{methodology} \citep{Azad2011}. During the JF season, 6 excess rainfall events were visible during the years 1984, 1990, 1994, 2000, 2011, and 2021. Deficient rainfall was not observed in this season. The MAM season revealed excess rainfall during 1972, 1995, 1999, 2004, 2006, 2015, and 2021 and deficient rainfall during 1979, 1983, 1986, 1987, and 2019. In the case of the JJAS season, both excess and deficient rainfall events were observable. 10 years of excess rainfall were recorded during the years 1968, 1975, 1981, 1991, 1992, 1994, 1997, 2007, 2013 and 2019. Similarly, 5 years of deficient rainfall were observed during the years 1976, 1987, 2002, 2012, and 2016. OND rainfall was excess during 1966, 1977, 1997, 2006, 2010, and 2021. Deficient rainfall was observed in 1967, 1974, 1988, 2012, and 2016.

The relative timing of solar activity and extreme rainfall was evaluated using curves of different solar indices, corresponding to different seasons (JF, MAM, JJAS, and OND). Figure \ref{ssnexcess}, \ref{f107excess} and \ref{criexcess} show the extreme rainfall events concerning SSN, F10.7, and CRI respectively. The black circle represents excess rainfall years and the red circle, the deficient rainfall years. The present study covers 5 complete solar cycles (Solar cycles 20-24). Tables \ref{ssnextremejf}, \ref{f10.7extremejf} and \ref{criextremejf} lists the extreme rainfall events during the JF season,  
Tables \ref{ssnextrememam}, \ref{f10.7extrememam} and \ref{criextrememam} during the MAM season, 
Tables \ref{ssnextremejjas}, \ref{f10.7extremejjas} and \ref{criextremejjas} during the JJAS season and Tables \ref{ssnextremeond}, \ref{f10.7extremeond} and \ref{criextremeond} during the OND season. The first column denotes the years of extreme, of each solar index (denoted as y), the second column, is the excess rainfall years, and the third column, is the deficient rainfall years. Values given in brackets indicate the relation with solar extremes. 

\subsubsection{Relation of extreme rainfall years with Sunspot number}

\begin{figure*}[t!]
\centering
  \begin{subfigure}{0.5\textwidth}
    \includegraphics[width=\linewidth]{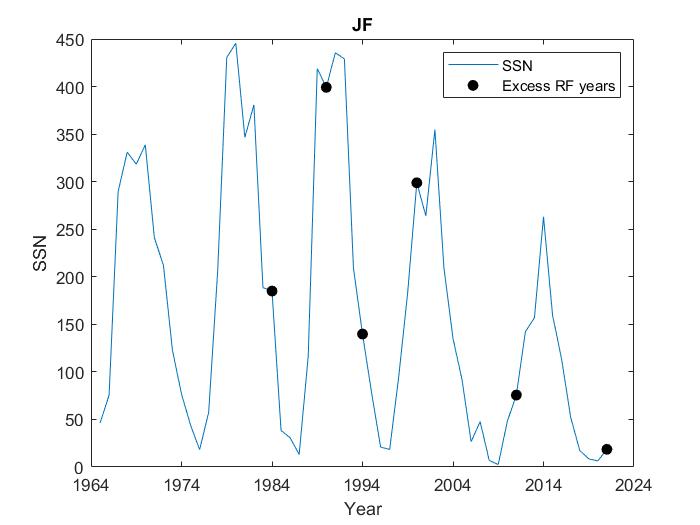}
    \caption{} \label{fig:13a}
  \end{subfigure}%
  \begin{subfigure}{0.5\textwidth}
    \includegraphics[width=\linewidth]{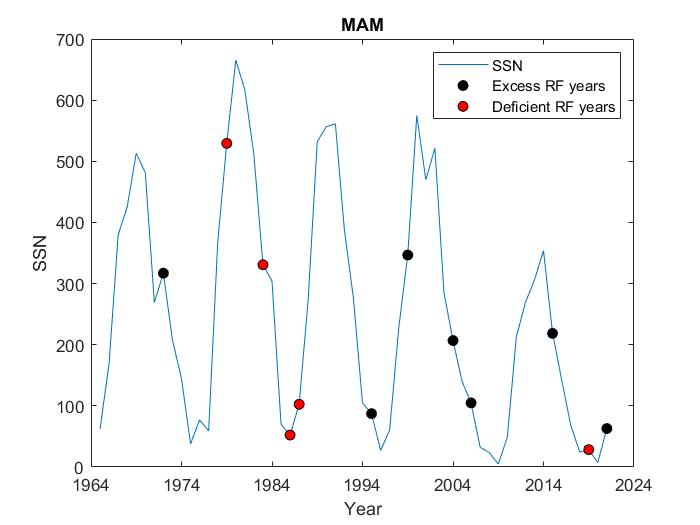}
    \caption{} \label{fig:13b}
  \end{subfigure}  \\
  \begin{subfigure}{0.5\textwidth}
    \includegraphics[width=\linewidth]{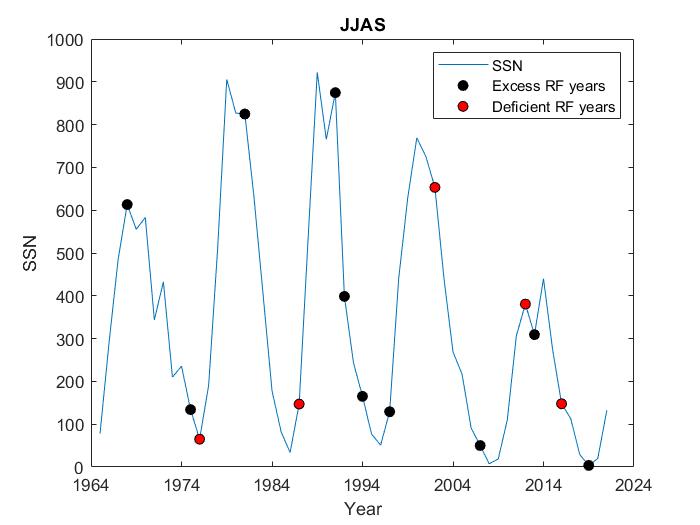}
    \caption{} \label{fig:13c}
  \end{subfigure}%
  \begin{subfigure}{0.5\textwidth}
    \includegraphics[width=\linewidth]{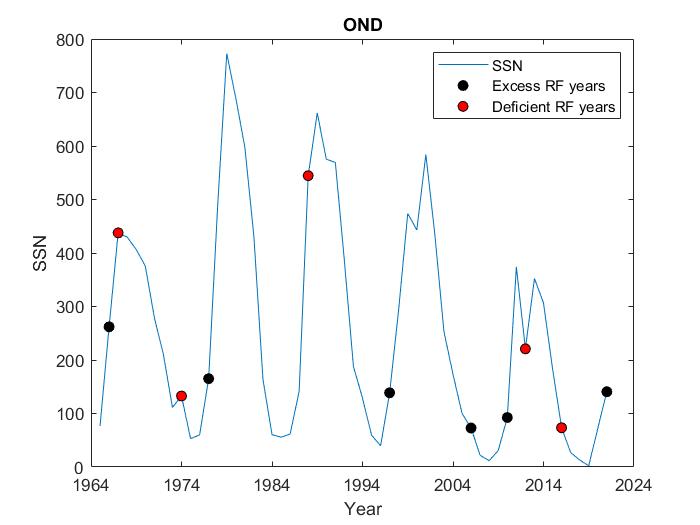}
    \caption{} \label{fig:13d}
  \end{subfigure}  \\
\caption{The extreme years of (a) JF (b) MAM (c) JJAS and (d) OND rainfall depicted on SSN plot}
\label{ssnexcess}
\end{figure*}

Figure \ref{ssnexcess} denotes the extreme rainfall years during the JF, MAM, JJAS, and OND seasons, plotted on the SSN curve. In this figure, slight variations in the maximum and minimum values of SSN during different seasons are visible. During the JF season, the SSN showed a maximum during the years 1970, 1980, 1991, 2002, and 2014 and a minimum during the years 1965, 1976, 1987, 1997, 2009, and 2020. Six excess rainfall years were observed during this season and are given in Table \ref{ssnextremejf}. 
The SSN values were maximum during 1969, 1980, 1991, 2000, and 2014, while they were minimum during 1965, 1975, 1986, 1996, 2009, and 2020 during the MAM season. Seven excess and five deficient rainfall years were identified and listed in Table \ref{ssnextrememam}.  
 In the JJAS season, maximum SSNs occurred in 1968, 1979, 1989, 2000, and 2014, and minimum SSNs occurred in 1965, 1976, 1986, 1996, 2008, and 2019. There were 10 excess and 5 deficient rainfall events noted during this season which are listed in Table \ref{ssnextremejjas}. In the case of the OND season, SSN values recorded maximums during 1967, 1979, 1989, 2001, and 2011, while minima were recorded during 1965, 1975, 1985, 1996, 2008, and 2019. Six excess rainfall years and five deficient rainfall years were noted and listed in Table \ref{ssnextremeond}.

 Considering all the seasons, it was commonly noted that the occurrences of extreme rainfall (excess or deficient) often occurred around 3 years of extreme solar activity, ie, a solar maximum/solar minimum. In the MAM season, a year of excess rainfall was noted to be 4 years after the solar maximum, which is an exception. During the JJAS and the MAM season, a few years of extreme rainfall events appeared to coincide with the years of solar extremes.  

\begin{table*}[!t]
   \centering
  \caption{Extreme rainfall years along with extreme SSN years, during JF season}
     \label{ssnextremejf}
    \begin{tabular}{cc} \hline
       Years of extreme SSN (y) & Excess rainfall years \\ \hline
    1965 (min) & \\
       1970 (max) & \\
       1976 (min) &  \\
       1980 (max) &  \\
       1987 (min) & 1984 (y-3) \\
       1991 (max) & 1990 (y-1) \\
       1997 (min) & 1994 (y-3)\\
       2002 (max) & 2000 (y-2) \\
       2009 (min) & 2011 (y+2) \\
       2014 (max) & \\
       2020 (min) & 2021 (y+1) \\ \hline
  \end{tabular}
   \end{table*}

\begin{table*}[!t]
   \centering
  \caption{Extreme rainfall years along with extreme SSN years, during MAM season}
     \label{ssnextrememam}
    \begin{tabular}{ccc} \hline
       Years of extreme SSN (y) & Excess rainfall years & Deficient rainfall years \\ \hline
       1965 (min) & & \\
       1969 (max) & & \\
       1975 (min) & 1972 (y-3) & \\
       1980 (max) & & 1979 (y-1) \\
       1986 (min) & & 1983 (y-3), 1986 (y), 1987 (y+1)  \\
       1991 (max) & &  \\
       1996 (min) & 1995 (y-1) & \\
       2000 (max) & 1999 (y-1), 2004 (y+4) & \\
       2009 (min) & 2006 (y-3) \\
       2014 (max) & 2015 (y+1) & \\
       2020 (min) & 2021 (y+1) & 2019 (y-1) \\ \hline
  \end{tabular}
   \end{table*}

\begin{table*}[!t]
   \centering
  \caption{Extreme rainfall years along with extreme SSN years, during JJAS season}
     \label{ssnextremejjas}
    \begin{tabular}{ccc} \hline
       Years of extreme SSN (y) & Excess rainfall years & Deficient rainfall years \\ \hline
       1965 (min) & & \\
       1968 (max) & 1968 (y) & \\
       1976 (min) & 1975 (y-1) & 1976 (y) \\
       1979 (max) & 1981 (y+2) & \\
       1986 (min) & &  1987 (y+1)  \\
       1989 (max) & 1991 (y+2), 1992 (y+3) &  \\
       1996 (min) & 1994 (y-2), 1997 (y+1) & \\
       2000 (max) &  & 2002 (y+2) \\
       2008 (min) & 2007 (y-1) \\
       2014 (max) & 2013 (y-1) & 2012 (y-2), 2016 (y+2)\\
       2019 (min) & 2019 (y) & \\ \hline
  \end{tabular}
   \end{table*}

\begin{table*}[!t]
   \centering
  \caption{Extreme rainfall years along with extreme SSN years, during OND season}
     \label{ssnextremeond}
    \begin{tabular}{ccc} \hline
       Years of extreme SSN (y) & Excess rainfall years & Deficient rainfall years \\ \hline
       1965 (min) & 1966 (y+1) & 1967 (y+2) \\
       1967 (max) &  & \\
       1975 (min) & 1977 (y+2) & 1974 (y-1) \\
       1979 (max) & & \\
       1985 (min) & &  \\
       1989 (max) & & 1988 (y-1) \\
       1996 (min) & 1997 (y+1) & \\
       2001 (max) &  &  \\
       2008 (min) & 2006 (y-2) \\
       2011 (max) & 2010 (y-1) & 2012 (y+1) \\
       2019 (min) & 2021 (y+3) & 2016 (y-3) \\ \hline
  \end{tabular}
   \end{table*}

\subsubsection{Relation of extreme rainfall years with F10.7 Index}
\begin{figure*}[t!]
\centering
  \begin{subfigure}{0.5\textwidth}
    \includegraphics[width=\linewidth]{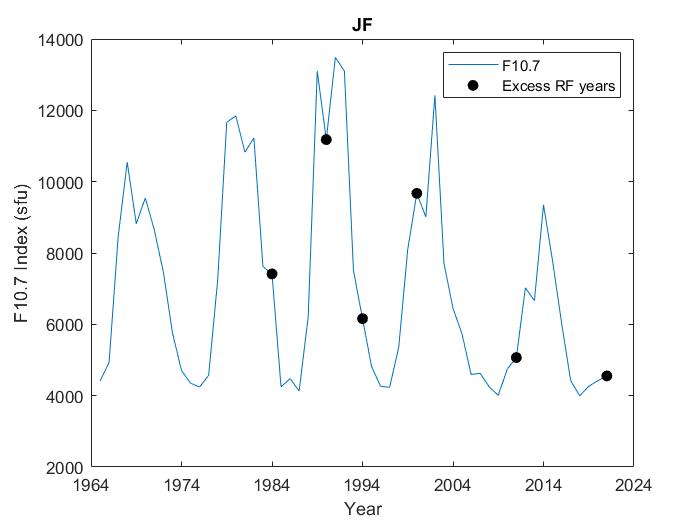}
    \caption{} \label{fig:14a}
  \end{subfigure}%
  \begin{subfigure}{0.5\textwidth}
    \includegraphics[width=\linewidth]{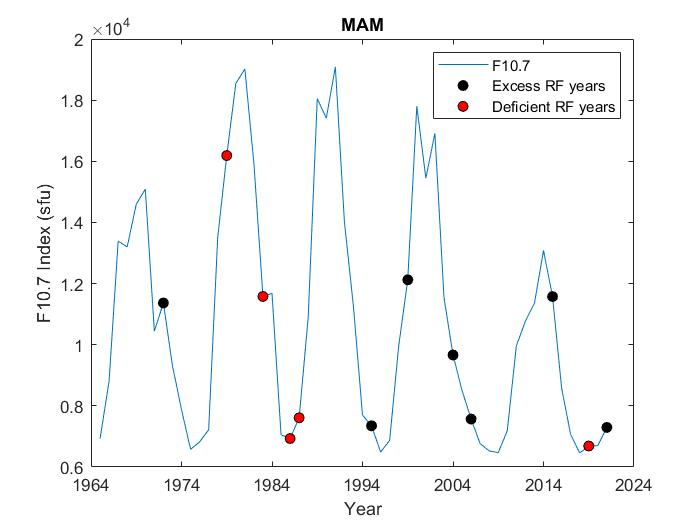}
    \caption{} \label{fig:14b}
  \end{subfigure}  \\
  \begin{subfigure}{0.5\textwidth}
    \includegraphics[width=\linewidth]{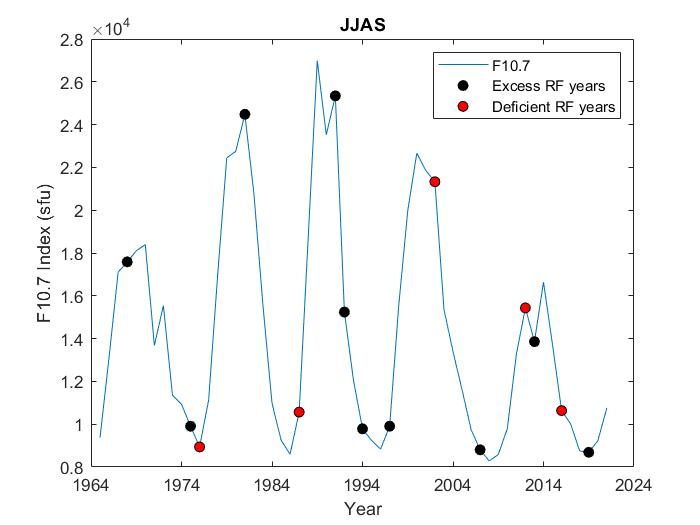}
    \caption{} \label{fig:14c}
  \end{subfigure}%
  \begin{subfigure}{0.5\textwidth}
    \includegraphics[width=\linewidth]{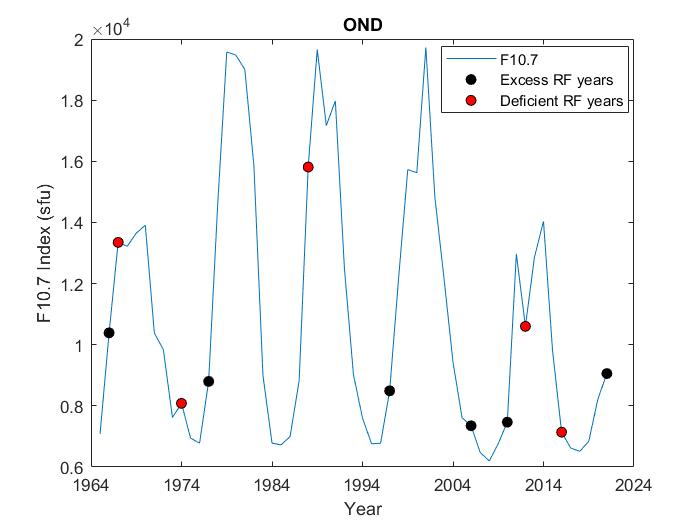}
    \caption{} \label{fig:14d}
  \end{subfigure}  \\
\caption{The extreme years of (a) JF (b) MAM (c) JJAS and (d) OND rainfall depicted on F10.7 plot}
\label{f107excess}
\end{figure*}

Similar to the SSN studies, we analyzed the extreme rainfall years in terms of the F10.7 Index. 
Figure \ref{f107excess} shows the extreme rainfall years during the JF, MAM, JJAS, and OND seasons, plotted on the F10.7 curve. Here, slight variations in the maximum and minimum values of F10.7, during different seasons, are again visible. During the JF season, the F10.7 showed a maximum during the years 1968, 1980, 1991, 2002, and 2014 and a minimum during the years 1965, 1976, 1987, 1997, 2009, and 2020. 
The F10.7 values were maximum during 1970, 1981, 1991, 2000, and 2014, while they were minimum during 1965, 1975, 1986, 1996, 2009, and 2018 during the MAM season. 
 In the JJAS season, maximum F10.7s occurred in 1970, 1981, 1989, 2000, and 2014, and minimum F10.7s occurred in 1965, 1976, 1986, 1996, 2008, and 2019. 
 In the case of the OND season, F10.7 values recorded maximums during 1970, 1979, 1989, 2001, and 2014, while minima were recorded during 1965, 1976, 1985, 1996, 2008, and 2018. The details of the extreme rainfall years about F10.7 Index are given in Tables \ref{f10.7extremejf}, \ref{f10.7extrememam}, \ref{f10.7extremejjas} and \ref{f10.7extremeond}.

 While considering F10.7 Index, results similar to that of the sunspot number were observed. Rainfall excesses and deficits occurred around three years of extreme solar index values, which are the times when solar activity is highest or lowest. The JJAS season analysis showed years when extreme rainfall and solar activity coincided.
 
  \begin{table*}[!t]
   \centering
  \caption{Extreme rainfall years along with extreme F10.7 years, during JF season}
     \label{f10.7extremejf}
    \begin{tabular}{cc} \hline
       Years of extreme F10.7 (y) & Excess rainfall years \\ \hline
       1965 (min) & \\
       1968 (max) & \\
       1976 (min) &  \\
       1980 (max) & \\
       1987 (min) & 1984 (y-3) \\
       1991 (max) & 1990 (y-1) \\
       1997 (min) & 1994 (y-3)\\
       2002 (max) & 2000 (y-2) \\
       2009 (min) & 2011 (y+2) \\
       2014 (max) & \\
       2018 (min) & 2021 (y+3) \\ \hline
  \end{tabular}
   \end{table*}

\begin{table*}[!t]
   \centering
  \caption{Extreme rainfall years along with extreme F10.7 years, during MAM season}
     \label{f10.7extrememam}
    \begin{tabular}{ccc} \hline
       Years of extreme F10.7 (y) & Excess rainfall years & Deficient rainfall years \\ \hline
       1965 (min) & & \\
       1970 (max) & 1972 (y+2) & \\
       1975 (min) & & \\
       1981 (max) & & 1979 (y-2) \\
       1986 (min) & & 1983 (y-3), 1986 (y), 1987 (y+1) \\
       1991 (max) &  & \\
       1996 (min) & 1995 (y-1) & \\
       2000 (max) &  & 1999 (y-1), 2004 (y+4) \\
       2009 (min) & 2006 (y-3) & \\
       2014 (max) & 2015 (y+1) &\\
       2018 (min) & 2021 (y+3) & 2019 (y+1) \\ \hline
  \end{tabular}
   \end{table*}

\begin{table*}[!t]
   \centering
  \caption{Extreme rainfall years along with extreme F10.7 years, during JJAS season}
     \label{f10.7extremejjas}
    \begin{tabular}{ccc} \hline
       Years of extreme F10.7 (y) & Excess rainfall years & Deficient rainfall years \\ \hline
       1965 (min) & & \\
       1970 (max) & 1968 (y-2) & \\
       1976 (min) & 1975 (y-1) & 1976 (y) \\
       1981 (max) & 1981 (y) & \\
       1986 (min) & & 1987 (y+1) \\
       1989 (max) & 1991 (y+2), 1992 (y+3) & \\
       1996 (min) & 1994 (y-2), 1997 (y+1) & \\
       2000 (max) &  & 2002 (y+2) \\
       2008 (min) & 2007 (y-1) \\
       2014 (max) & 2013 (y-1) & 2012 (y-2), 2016 (y+2)\\
       2019 (min) & 2019 (y) & \\ \hline
  \end{tabular}
   \end{table*}

\begin{table*}[!t]
   \centering
  \caption{Extreme rainfall years along with extreme F10.7 years, during OND season}
     \label{f10.7extremeond}
    \begin{tabular}{ccc} \hline
       Years of extreme F10.7 (y) & Excess rainfall years & Deficient rainfall years \\ \hline
       1965 (min) & 1966 (y+1) & 1967 (y+2)\\
       1970 (max) &  & \\
       1976 (min) & 1977 (y+1) & 1974 (y-2) \\
       1979 (max) & & \\
       1985 (min) & &  \\
       1989 (max) & & 1988 (y-1) \\
       1996 (min) & 1997 (y+1) & \\
       2001 (max) &  & \\
       2008 (min) & 2006 (y-2) \\
       2014 (max) & 2010 (y-4) & 2012 (y-2) \\
       2018 (min) & 2021 (y+3) & 2016 (y-2) \\ \hline
  \end{tabular}
   \end{table*}

\subsubsection{Relation of extreme rainfall years with Cosmic ray intensity}

\begin{figure*}[t!]
\centering
  \begin{subfigure}{0.5\textwidth}
    \includegraphics[width=\linewidth]{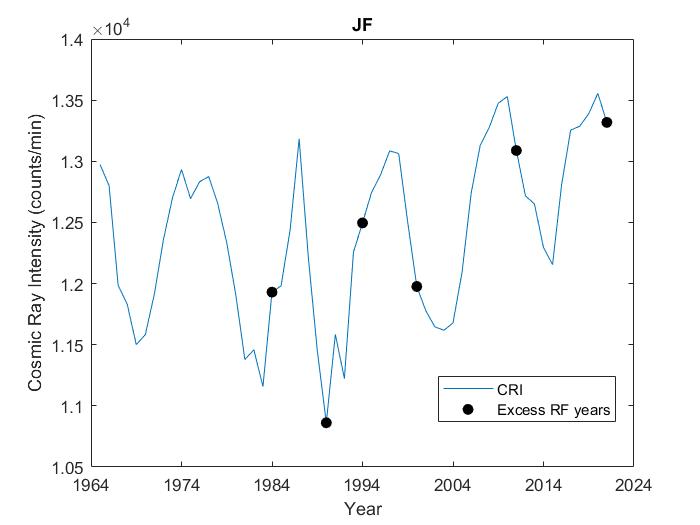}
    \caption{} \label{fig:15a}
  \end{subfigure}%
  \begin{subfigure}{0.5\textwidth}
    \includegraphics[width=\linewidth]{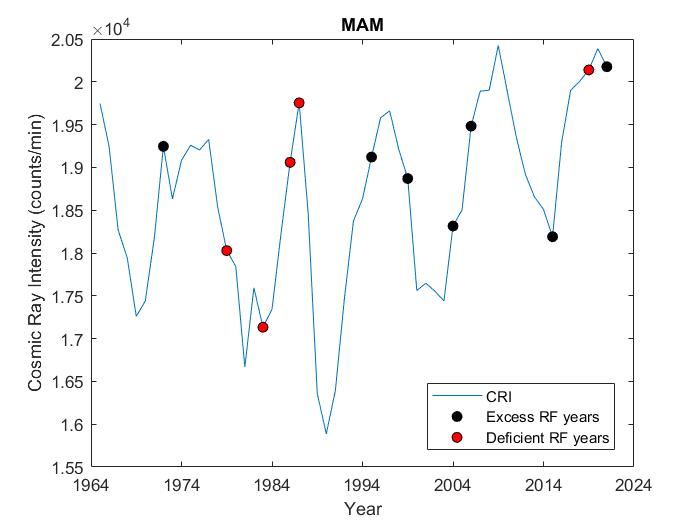}
    \caption{} \label{fig:15b}
  \end{subfigure}  \\
  \begin{subfigure}{0.5\textwidth}
    \includegraphics[width=\linewidth]{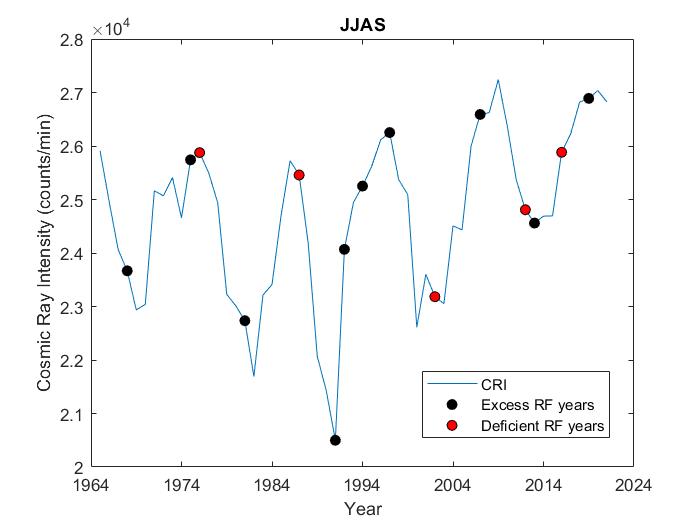}
    \caption{} \label{fig:15c}
  \end{subfigure}%
  \begin{subfigure}{0.5\textwidth}
    \includegraphics[width=\linewidth]{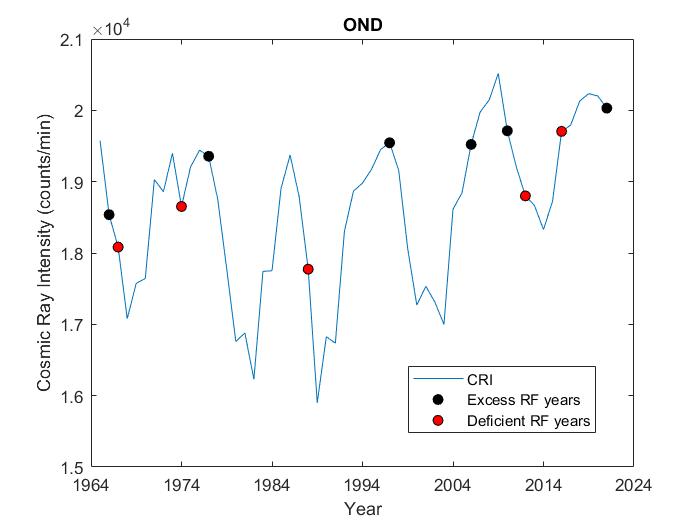}
    \caption{} \label{fig:15d}
  \end{subfigure}  \\
\caption{The extreme years of (a) JF (b) MAM (c) JJAS and (d) OND rainfall depicted on CRI plot}
\label{criexcess}
\end{figure*}

Lastly, the relationship of extreme rainfall years with CRI was studied. Figure \ref{criexcess} depicts the extreme rainfall years during the different seasons plotted on CRI curves. During the JF season, CRI displayed maximum values during the years 1965, 1974, 1987, 1997, 2010, and 2020, and minimum values during the years 1969, 1983, 1990, 2003, and 2015. The CRI values were maximum during 1965, 1977, 1987, 1997, 2009, and 2015, while they were minimum during 1969, 1981, 1990, 2003, and 2015 during the MAM season.
In the JJAS season, CRI showed maximum values in 1965, 1976, 1986, 1997, 2009, and 2020 and minimum values in 1969, 1982, 1991, 2000, and 2015. During the OND season, CRI displayed maximum values during the years 1965, 1976, 1986, 1997, 2009, and 2019, and minimum values during the years 1968, 1982, 1989, 2003, and 2014. List of the extreme rainfall years during the JF, MAM, JJAS, and OND seasons is given in Tables \ref{criextremejf}, \ref{criextrememam}, \ref{criextremejjas} and \ref{criextremeond} respectively. Comparing with the previous results of sunspot number and F10.7 Index, cosmic ray intensity also showed extreme values around the years of extreme rainfall events. MAM and JJAS seasons revealed years where the two events were coinciding. 

\begin{table*}[!t]
   \centering
  \caption{Extreme rainfall years along with extreme CRI years, during JF season}
     \label{criextremejf}
    \begin{tabular}{cc} \hline
       Years of extreme CRI (y) & Excess rainfall years \\ \hline
       1965 (min) & \\
       1969 (max) & \\
       1974 (min) &  \\
       1983 (max) & 1984 (y+1) \\
       1987 (min) &  \\
       1990 (max) & 1990 (y) \\
       1997 (min) & 1994 (y-3) \\
       2003 (max) & 2000 (y-3) \\
       2010 (min) & 2011 (y+1) \\
       2015 (max) & \\
       2020 (min) & 2021 (y+1) \\ \hline
  \end{tabular}
   \end{table*}

\begin{table*}[!t]
   \centering
  \caption{Extreme rainfall years along with extreme CRI years, during MAM season}
     \label{criextrememam}
    \begin{tabular}{ccc} \hline
       Years of extreme CRI (y) & Excess rainfall years & Deficient rainfall years \\ \hline
       1965 (min) & & \\
       1969 (max) & 1972 (y+3) & \\
       1977 (min) &  &  \\
       1981 (max) &  & 1979 (y-2), 1983 (y+2) \\
       1987 (min) & & 1986 (y-1), 1987 (y+1)\\
       1990 (max) &  & \\
       1997 (min) & 1995 (y-2), 1999 (y+2) & \\
       2003 (max) & 2004 (y+1), 2006 (y+3) &  \\
       2009 (min) &  & \\
       2015 (max) & 2015 (y) & 2016 \\
       2020 (min) & 2021 (y+1) & 2019 (y-1)\\ \hline
  \end{tabular}
   \end{table*}

\begin{table*}[!t]
   \centering
  \caption{Extreme rainfall years along with extreme CRI years, during JJAS season}
     \label{criextremejjas}
    \begin{tabular}{ccc} \hline
       Years of extreme CRI (y) & Excess rainfall years & Deficient rainfall years \\ \hline
       1965 (min) & & \\
       1969 (max) & 1968 (y-1) & \\
       1976 (min) & 1975 (y-1) &  1976 (y) \\
       1982 (max) & 1981 (y-1) & \\
       1986 (min) & & 1987 (y+1)\\
       1991 (max) & 1991 (y), 1992 (y+1), 1993 (y+3) & \\
       1997 (min) & 1997 (y) & \\
       2000 (max) &  & 2002 (y+2) \\
       2009 (min) & 2007 (y-2)  &\\
       2015 (max) & 2013 (y-2) & 2012 (y-3), 2016 (y+1)\\
       2020 (min) & 2019 (y-1) & \\ \hline
  \end{tabular}
   \end{table*}

   \begin{table*}[!t]
   \centering
  \caption{Extreme rainfall years along with extreme CRI years, during OND season}
     \label{criextremeond}
    \begin{tabular}{ccc} \hline
       Years of extreme CRI (y) & Excess rainfall years & Deficient rainfall years \\ \hline
       1965 (min) & 1966 (y+1) & 1967 (y+2)\\
       1968 (max) & & \\
       1976 (min) & 1977 (y+1) &  1974 (y-2) \\
       1982 (max) &  & \\
       1986 (min) & & \\
       1989 (max) &  & 1988 (y-1)\\
       1997 (min) & 1997 (y) & \\
       2003 (max) &  &  \\
       2009 (min) & 2006 (y-3)  & \\
       2014 (max) & 2010 (y-4) & 2012 (y-2)\\
       2019 (min) & 2021 (y+2) & 2016 (y-3)\\ \hline
  \end{tabular}
   \end{table*}

According to the findings of the study, the results are in agreement with those obtained in earlier Indian studies. \cite{bhalme1981cyclic} observed that Flood Area Index over India was associated with the double sunspot cycle. 
During alternate solar cycles, \cite{Ananthakrishnan1984} noted significantly more excess rainfall years during the ascending phase. 
According to \cite{jain1997correlation}, the periodicity of floods and droughts are well correlated with sunspot main periods and quasi-periods in the Udaipur subtropical region of Rajasthan. \cite{Bhattacharyya2005} have shown that high rainfall correlates with high solar activity, while low rainfall correlates with low solar activity. Considering the sub-divisions from west central and peninsular India, \cite{Azad2011} reported that the maxima of even sunspot cycles coincided with excess rainfall (with +1 year error) and the minima of odd sunspot cycles coincided with deficit rainfall (with $\pm$2 year error). 

Globally, studies have examined how solar activity impacts extreme rainfall events. \cite{mitchell1979,cook1997} observed that the drought cycle is related to the double (Hale) sunspot cycle in the United States. \cite{Vaquero2004SolarSI} noted that the probability of floods increased during the episodes of high solar activity while studying the floods at the Tagus river basin, Central Spain. A study on Lake Victoria levels in East Africa showed that solar activity affects levels through rainfall. Rainfall maxima lagged one year behind sunspot maxima, resulting in lake level maxima \citep{Stager2007}. Sunspot number revealed a direct correlation with the flood/drought of the Second Songhua river basin, China \citep{hong2015}.  In typical regions of the Loess Plateau, in Yan'an, China, studies of precipitation responses to the solar activity found that maximum precipitation was observed during solar maximum and was associated with solar activity \citep{Li2017}. \cite{YU2019} observed that the occurrences of droughts and floods in the Southern Chinese Loess Plateau were synchronous with solar activities, at least on decadal timescales. 

In some cases, contradictory results were reported.  \cite{Wirth2013} noted that flood frequency in the European Alps increased during cool periods, which coincided with low solar activity. The frequency of flood years is relatively high when solar activity is low, and vice versa, according to \cite{Rimbu2021} studies on River Ammer floods in Germany.  \cite{Li2023} studied the time-lag correlation between solar activity and summer precipitation in the mid-lower reaches of the Yangtze River, China. It was noted that the sunspot number has a negative correlation with precipitation, with an 11-month time lag.

Concerns about the solar impact on rainfall have been around for a while, and numerous connections that may exist have been examined in various studies \citep{Li2023}. Similar periods in the time series of rainfall and solar activity suggested a potential relationship between them \citep{Nitka2019, HEREDIA2019105094}. Sea surface temperature can be influenced by the total solar irradiance (TSI), which alters atmospheric circulation and modulates rainfall \citep{Soon1996}. A temperature gradient results from the sun's UV rays being absorbed by the stratospheric ozone. Through interactions between the troposphere and stratosphere, this in turn influences Brewer-Dobson circulation and further changes the lower atmosphere \citep{Baldwin2005}. The production of cloud condensation nuclei \citep{Svensmark2007} and, eventually, precipitation are influenced by galactic cosmic rays. 

\section{Conclusion}\label{conclusions}
The association between solar activity and the extreme precipitation events over Kerala, India, was examined in this study. The fluctuation of the three solar indices—sunspot number, F10.7 Index, and cosmic ray intensity—with rainfall was examined over a 57-year period (1965–2021). By separating the solar and rainfall data into the winter, pre-monsoon, monsoon, and post-monsoon seasons, both annual and seasonal scales were taken into consideration. The correlation coefficients between seasonal rainfall and the solar indices were computed. The rainfall in Kerala correlated with the sunspot activity, however the association had varied significance, according to correlation analysis. When solar activity was high, there was a strong association between the winter and monsoon seasons and their significance. 
Sunspot number and F10.7 index showed stronger correlation than cosmic ray intensity among the three solar indices taken into account. In contrast to other areas, Kerala's central and southern portions appear to be affected by the sun at times of high activity.
The years when Kerala saw excessive or insufficient rainfall were identified, and its relationship with the various solar indices was investigated. 
When the solar activity is at its peak, or around solar maximums and minimums, excess and deficit rainfall years tend to occur. 
This study raises the possibility of a physical link, which increases predictability, between solar activity and extreme precipitation in Kerala. 

\section*{Data Availability}
The dataset on sunspot number is available at 
\url{http://www.sidc.be/silso/data files}. The F10.7 Index data is available at \url{https://lasp.colorado.edu/lisird/data/cls_radio_flux_f107}. Cosmic ray intensity data is available at \url{https://cosmicrays.oulu.fi}. Gridded rainfall data is available at \url{https://www.imdpune.gov.in/lrfindex.php}.

\section*{CRediT authorship contribution statement}
Elizabeth Thomas: Conceptualization, Methodology, Formal analysis and investigation, Data curation, Writing - original draft preparation. S. Vineeth: Formal analysis and investigation, Data curation. Noble P. Abraham: Conceptualization, Writing - review and editing, Supervision.

\section*{Declaration of Competing Interest}
The authors declare that they have no known competing financial
interests or personal relationships that could have appeared to influence the work reported in this paper.

\section*{Acknowledgement}
 First author acknowledges the financial assistance from the University Grants Commission (UGC), India, under Savitribai Jyotirao Phule Fellowship for Single Girl Child (SJSGC) (F. No. 82-7/2022(SA-III) dated 07/02/2023). Second author acknowledges the financial assistance from Department of Science and Technology (DST), Ministry of Science and Technology, India under INSPIRE Fellowship (Award Letter No. IF180235 dated 08/02/2019).




\bibliographystyle{elsarticle-harv} 
\bibliography{example}

\begin{thebibliography}{63}
\expandafter\ifx\csname natexlab\endcsname\relax\def\natexlab#1{#1}\fi
\providecommand{\url}[1]{\texttt{#1}}
\providecommand{\href}[2]{#2}
\providecommand{\path}[1]{#1}
\providecommand{\DOIprefix}{doi:}
\providecommand{\ArXivprefix}{arXiv:}
\providecommand{\URLprefix}{URL: }
\providecommand{\Pubmedprefix}{pmid:}
\providecommand{\doi}[1]{\href{http://dx.doi.org/#1}{\path{#1}}}
\providecommand{\Pubmed}[1]{\href{pmid:#1}{\path{#1}}}
\providecommand{\bibinfo}[2]{#2}
\ifx\xfnm\relax \def\xfnm[#1]{\unskip,\space#1}\fi
\bibitem[{Agnihotri et~al.(2011)Agnihotri, Dutta and Soon}]{Agnihotri2011}
\bibinfo{author}{Agnihotri, R.}, \bibinfo{author}{Dutta, K.}, \bibinfo{author}{Soon, W.}, \bibinfo{year}{2011}.
\newblock \bibinfo{title}{{Temporal derivative of Total Solar Irradiance and anomalous Indian summer monsoon: An empirical evidence for a Sun-climate connection}}.
\newblock \bibinfo{journal}{J. Atmos. Sol.-Terr. Phys.} \bibinfo{volume}{73}, \bibinfo{pages}{1980--1987}.
\newblock \DOIprefix\doi{10.1016/j.jastp.2011.06.006}.
\bibitem[{Ananthakrishnan and Parthasarathy(1984)}]{Ananthakrishnan1984}
\bibinfo{author}{Ananthakrishnan, R.}, \bibinfo{author}{Parthasarathy, B.}, \bibinfo{year}{1984}.
\newblock \bibinfo{title}{{Indian rainfall in relation to the sunspot cycle: 1871-1978}}.
\newblock \bibinfo{journal}{J. Climatol.} \bibinfo{volume}{4}, \bibinfo{pages}{149--169}.
\newblock \DOIprefix\doi{10.1002/joc.3370040205}.
\bibitem[{Azad(2011)}]{Azad2011}
\bibinfo{author}{Azad, S.}, \bibinfo{year}{2011}.
\newblock \bibinfo{title}{{Extreme Indian Monsoon Rainfall Years and the Sunspot Cycle}}.
\newblock \bibinfo{journal}{Adv. Sci. Lett.} \bibinfo{volume}{4}, \bibinfo{pages}{159--164}.
\newblock \DOIprefix\doi{10.1166/asl.2011.1203}.
\bibitem[{Badruddin and Aslam(2015)}]{Badruddin2015}
\bibinfo{author}{Badruddin}, \bibinfo{author}{Aslam, O.P.}, \bibinfo{year}{2015}.
\newblock \bibinfo{title}{{Influence of cosmic-ray variability on the monsoon rainfall and temperature}}.
\newblock \bibinfo{journal}{J. Atmos. Sol.-Terr. Phys.} \bibinfo{volume}{122}, \bibinfo{pages}{86--96}.
\newblock \DOIprefix\doi{10.1016/j.jastp.2014.11.005}, \href{http://arxiv.org/abs/1412.1041}{{\tt arXiv:1412.1041}}.
\bibitem[{Bal and Bose(2010)}]{Bal2010}
\bibinfo{author}{Bal, S.}, \bibinfo{author}{Bose, M.}, \bibinfo{year}{2010}.
\newblock \bibinfo{title}{{A climatological study of the relations among solar activity, galactic cosmic ray and precipitation on various regions over the globe}}.
\newblock \bibinfo{journal}{J. Earth Syst. Sci.} \bibinfo{volume}{119}, \bibinfo{pages}{201--209}.
\newblock \DOIprefix\doi{10.1007/s12040-010-0015-8}.
\bibitem[{Baldwin and Dunkerton(2005)}]{Baldwin2005}
\bibinfo{author}{Baldwin, M.P.}, \bibinfo{author}{Dunkerton, T.J.}, \bibinfo{year}{2005}.
\newblock \bibinfo{title}{{The solar cycle and stratosphere-troposphere dynamical coupling}}.
\newblock \bibinfo{journal}{J. Atmos. Sol. Terr. Phys.} \bibinfo{volume}{67}, \bibinfo{pages}{71--82}.
\newblock \DOIprefix\doi{10.1016/j.jastp.2004.07.018}.
\bibitem[{Bankoti et~al.(2011)Bankoti, Joshi, Pande, Pande and Pandey}]{Bankoti2011}
\bibinfo{author}{Bankoti, N.S.}, \bibinfo{author}{Joshi, N.C.}, \bibinfo{author}{Pande, S.}, \bibinfo{author}{Pande, B.}, \bibinfo{author}{Pandey, K.}, \bibinfo{year}{2011}.
\newblock \bibinfo{title}{{Correlative study of different solar activity features with all India homogeneous rainfall during 1963-2006}}.
\newblock \bibinfo{journal}{Quat. Int} \bibinfo{volume}{229}, \bibinfo{pages}{8--15}.
\newblock \DOIprefix\doi{10.1016/j.quaint.2010.04.006}.
\bibitem[{Bhalme and Mooley(1981)}]{bhalme1981cyclic}
\bibinfo{author}{Bhalme, H.}, \bibinfo{author}{Mooley, D.}, \bibinfo{year}{1981}.
\newblock \bibinfo{title}{Cyclic fluctuations in the flood area and relationship with the double (hale) sunspot cycle}.
\newblock \bibinfo{journal}{J. Appl. Meteorol.(1962-1982)} , \bibinfo{pages}{1041--1048}.
\bibitem[{Bhalme et~al.(1981)Bhalme, Reddy, Mooley and Murty}]{bhalme1981solar}
\bibinfo{author}{Bhalme, H.}, \bibinfo{author}{Reddy, R.}, \bibinfo{author}{Mooley, D.}, \bibinfo{author}{Murty, B.V.R.}, \bibinfo{year}{1981}.
\newblock \bibinfo{title}{Solar activity and indian weather/climate}.
\newblock \bibinfo{journal}{Proc. Indian Acad, Sci. (Earth Planet. Sci.)} \bibinfo{volume}{90}, \bibinfo{pages}{245--262}.
\bibitem[{Bhattacharyya and Narasimha(2005)}]{Bhattacharyya2005}
\bibinfo{author}{Bhattacharyya, S.}, \bibinfo{author}{Narasimha, R.}, \bibinfo{year}{2005}.
\newblock \bibinfo{title}{{Possible association between Indian monsoon rainfall and solar activity}}.
\newblock \bibinfo{journal}{Geophysical Research Letters} \bibinfo{volume}{32}, \bibinfo{pages}{1--5}.
\newblock \DOIprefix\doi{10.1029/2004GL021044}.
\bibitem[{Bhattacharyya and Narasimha(2007)}]{Bhattacharyya2007}
\bibinfo{author}{Bhattacharyya, S.}, \bibinfo{author}{Narasimha, R.}, \bibinfo{year}{2007}.
\newblock \bibinfo{title}{{Regional differentiation in multidecadal connections between Indian monsoon rainfall and solar activity}}.
\newblock \bibinfo{journal}{J. Geophys. Res. Atmos.} \bibinfo{volume}{112}, \bibinfo{pages}{1--10}.
\newblock \DOIprefix\doi{10.1029/2006JD008353}.
\bibitem[{Chakraborty and Bondyopadhyay(1986)}]{chakraborty1986solar}
\bibinfo{author}{Chakraborty, P.}, \bibinfo{author}{Bondyopadhyay, R.}, \bibinfo{year}{1986}.
\newblock \bibinfo{title}{Solar effect on rainfall in west bengal}.
\newblock \bibinfo{journal}{Mausam} \bibinfo{volume}{37}, \bibinfo{pages}{251--258}.
\bibitem[{Chaudhuri et~al.(2015)Chaudhuri, Pal and Guhathakurta}]{Chaudhuri2015}
\bibinfo{author}{Chaudhuri, S.}, \bibinfo{author}{Pal, J.}, \bibinfo{author}{Guhathakurta, S.}, \bibinfo{year}{2015}.
\newblock \bibinfo{title}{{The influence of galactic cosmic ray on all India annual rainfall and temperature}}.
\newblock \bibinfo{journal}{Adv. Space Res.} \bibinfo{volume}{55}, \bibinfo{pages}{1158--1167}.
\newblock \DOIprefix\doi{10.1016/j.asr.2014.11.027}.
\bibitem[{Cook et~al.(1997)Cook, Meko and Stockton}]{cook1997}
\bibinfo{author}{Cook, E.R.}, \bibinfo{author}{Meko, D.M.}, \bibinfo{author}{Stockton, C.W.}, \bibinfo{year}{1997}.
\newblock \bibinfo{title}{A new assessment of possible solar and lunar forcing of the bidecadal drought rhythm in the western united states}.
\newblock \bibinfo{journal}{J. Clim.} \bibinfo{volume}{10}, \bibinfo{pages}{1343 -- 1356}.
\newblock \DOIprefix\doi{10.1175/1520-0442(1997)010<1343:ANAOPS>2.0.CO;2}.
\bibitem[{{Doranalu Chandrashekar} et~al.(2017){Doranalu Chandrashekar}, Shetty, Singh and Sharma}]{DoranaluChandrashekar2017}
\bibinfo{author}{{Doranalu Chandrashekar}, V.}, \bibinfo{author}{Shetty, A.}, \bibinfo{author}{Singh, B.B.}, \bibinfo{author}{Sharma, S.}, \bibinfo{year}{2017}.
\newblock \bibinfo{title}{{Spatio-temporal precipitation variability over Western Ghats and Coastal region of Karnataka, envisaged using high resolution observed gridded data}}.
\newblock \bibinfo{journal}{Model. Earth Syst. Environ.} \bibinfo{volume}{3}, \bibinfo{pages}{1611--1625}.
\newblock \DOIprefix\doi{10.1007/s40808-017-0395-8}.
\bibitem[{Gupta et~al.(2006)Gupta, Mishra and Mishra}]{Gupta2006}
\bibinfo{author}{Gupta, M.}, \bibinfo{author}{Mishra, V.K.}, \bibinfo{author}{Mishra, A.P.}, \bibinfo{year}{2006}.
\newblock \bibinfo{title}{{Correlation of the long-term cosmic ray intensity variations with sunspot numbers and tilt angle}}.
\newblock \bibinfo{journal}{Indian J. Radio Space Phys} \bibinfo{volume}{35}, \bibinfo{pages}{387--395}.
\bibitem[{Hathaway(2015)}]{hathaway2015solar}
\bibinfo{author}{Hathaway, D.H.}, \bibinfo{year}{2015}.
\newblock \bibinfo{title}{The solar cycle}.
\newblock \bibinfo{journal}{Living Rev. Sol. Phys.} \bibinfo{volume}{12}, \bibinfo{pages}{4}.
\newblock \DOIprefix\doi{10.1007/lrsp-2015-4}.
\bibitem[{Heredia et~al.(2019)Heredia, Bazzano, Cionco, Soon, Medina and Elias}]{HEREDIA2019105094}
\bibinfo{author}{Heredia, T.}, \bibinfo{author}{Bazzano, F.M.}, \bibinfo{author}{Cionco, R.G.}, \bibinfo{author}{Soon, W.}, \bibinfo{author}{Medina, F.D.}, \bibinfo{author}{Elias, A.G.}, \bibinfo{year}{2019}.
\newblock \bibinfo{title}{Searching for solar-like interannual to bidecadal effects on temperature and precipitation over a southern hemisphere location}.
\newblock \bibinfo{journal}{J. Atmos. Sol.-Terr. Phys} \bibinfo{volume}{193}, \bibinfo{pages}{105094}.
\newblock \DOIprefix\doi{https://doi.org/10.1016/j.jastp.2019.105094}.
\bibitem[{Hiremath(2006)}]{Hiremath2006}
\bibinfo{author}{Hiremath, K.M.}, \bibinfo{year}{2006}.
\newblock \bibinfo{title}{{The Influence of Solar Activity on the Rainfall over India: Cycle-to-Cycle Variations}}.
\newblock \bibinfo{journal}{J. Astrophys. Astr.} \bibinfo{volume}{27}, \bibinfo{pages}{367--372}.
\bibitem[{Hiremath and Mandi(2004)}]{Hiremath2004}
\bibinfo{author}{Hiremath, K.M.}, \bibinfo{author}{Mandi, P.I.}, \bibinfo{year}{2004}.
\newblock \bibinfo{title}{{Influence of the solar activity on the Indian Monsoon rainfall}}.
\newblock \bibinfo{journal}{New Astron.} \bibinfo{volume}{9}, \bibinfo{pages}{651--662}.
\newblock \DOIprefix\doi{10.1016/j.newast.2004.04.001}.
\bibitem[{Hiremath et~al.(2015)Hiremath, Manjunath and Soon}]{Hiremath2015}
\bibinfo{author}{Hiremath, K.M.}, \bibinfo{author}{Manjunath, H.}, \bibinfo{author}{Soon, W.}, \bibinfo{year}{2015}.
\newblock \bibinfo{title}{{Indian summer monsoon rainfall: Dancing with the tunes of the sun}}.
\newblock \bibinfo{journal}{New Astron.} \bibinfo{volume}{35}, \bibinfo{pages}{8--19}.
\newblock \DOIprefix\doi{10.1016/j.newast.2014.08.002}.
\bibitem[{Hong-yan et~al.(2015)Hong-yan, Li-jun and Wang}]{hong2015}
\bibinfo{author}{Hong-yan, L.}, \bibinfo{author}{Li-jun, X.}, \bibinfo{author}{Wang, X.}, \bibinfo{year}{2015}.
\newblock \bibinfo{title}{Relationship between solar activity and flood/drought disasters of the second songhua river basin}.
\newblock \bibinfo{journal}{J. Water Clim. Chang.} \bibinfo{volume}{6}, \bibinfo{pages}{578}.
\newblock \DOIprefix\doi{10.2166/wcc.2014.053}.
\bibitem[{Jagannathan and Bhalme(1973)}]{jagannathan1973changes}
\bibinfo{author}{Jagannathan, P.}, \bibinfo{author}{Bhalme, H.}, \bibinfo{year}{1973}.
\newblock \bibinfo{title}{Changes in the pattern of distribution of southwest monsoon rainfall over india associated with sunspots}.
\newblock \bibinfo{journal}{Mon. Weather Rev.} \bibinfo{volume}{101}, \bibinfo{pages}{691--700}.
\newblock \DOIprefix\doi{10.1175/1520-0493(1973)101<0691:citpod>2.3.co;2}.
\bibitem[{Jagannathan and Parthasarathy(1973)}]{JAGANNATHAN1973}
\bibinfo{author}{Jagannathan, P.}, \bibinfo{author}{Parthasarathy, B.}, \bibinfo{year}{1973}.
\newblock \bibinfo{title}{{Trends and Periodicities of Rainfall Over India}}.
\newblock \bibinfo{journal}{Mon. Weather Rev.} \bibinfo{volume}{101}, \bibinfo{pages}{371--375}.
\newblock \DOIprefix\doi{10.1175/1520-0493(1973)101<0371:taporo>2.3.co;2}.
\bibitem[{Jain and Tripathy(1997)}]{jain1997correlation}
\bibinfo{author}{Jain, R.}, \bibinfo{author}{Tripathy, S.C.}, \bibinfo{year}{1997}.
\newblock \bibinfo{title}{{Correlation study between sunspot and rainfall in Udaipur subregion}}.
\newblock \bibinfo{journal}{Mausam} \bibinfo{volume}{48}, \bibinfo{pages}{405--412}.
\bibitem[{Kothawale and Rajeevan(2017)}]{Kothawale2017}
\bibinfo{author}{Kothawale, D.R.}, \bibinfo{author}{Rajeevan, M.}, \bibinfo{year}{2017}.
\newblock \bibinfo{title}{{Monthly , Seasonal and Annual Rainfall Time Series for All-India , Homogeneous Regions and Meteorological Subdivisions : 1871-2016}}.
\newblock \bibinfo{journal}{Indian Institute of Tropical Meteorology (IITM) Earth System Science Organization, Ministry of Earth Sciences} \bibinfo{volume}{02}, \bibinfo{pages}{1--164}.
\bibitem[{Krishnakumar et~al.(2009)Krishnakumar, {Prasada Rao} and Gopakumar}]{Krishnakumar2009}
\bibinfo{author}{Krishnakumar, K.N.}, \bibinfo{author}{{Prasada Rao}, G.S.}, \bibinfo{author}{Gopakumar, C.S.}, \bibinfo{year}{2009}.
\newblock \bibinfo{title}{{Rainfall trends in twentieth century over Kerala, India}}.
\newblock \bibinfo{journal}{Atmos. Environ} \bibinfo{volume}{43}, \bibinfo{pages}{1940--1944}.
\newblock \DOIprefix\doi{10.1016/j.atmosenv.2008.12.053}.
\bibitem[{Laurenz et~al.(2019)Laurenz, L{\"{u}}decke and L{\"{u}}ning}]{Laurenz2019}
\bibinfo{author}{Laurenz, L.}, \bibinfo{author}{L{\"{u}}decke, H.J.}, \bibinfo{author}{L{\"{u}}ning, S.}, \bibinfo{year}{2019}.
\newblock \bibinfo{title}{{Influence of solar activity changes on European rainfall}}.
\newblock \bibinfo{journal}{J. Atmos. Sol.-Terr. Phys.} \bibinfo{volume}{185}, \bibinfo{pages}{29--42}.
\newblock \DOIprefix\doi{10.1016/j.jastp.2019.01.012}.
\bibitem[{Li et~al.(2023)Li, Wang and Wang}]{Li2023}
\bibinfo{author}{Li, H.}, \bibinfo{author}{Wang, Y.}, \bibinfo{author}{Wang, C.}, \bibinfo{year}{2023}.
\newblock \bibinfo{title}{{Lagged response of summer precipitation to solar activity in the mid-lower reaches of the Yangtze River}}.
\newblock \bibinfo{journal}{Front. Earth Sci.} \bibinfo{volume}{10}, \bibinfo{pages}{1--10}.
\newblock \DOIprefix\doi{10.3389/feart.2022.1101252}.
\bibitem[{Li et~al.(2017)Li, Gao, Zhang, Zhang and Zhang}]{Li2017}
\bibinfo{author}{Li, H.J.}, \bibinfo{author}{Gao, J.E.}, \bibinfo{author}{Zhang, H.C.}, \bibinfo{author}{Zhang, Y.X.}, \bibinfo{author}{Zhang, Y.Y.}, \bibinfo{year}{2017}.
\newblock \bibinfo{title}{{Response of Extreme Precipitation to Solar Activity and El Nino Events in Typical Regions of the Loess Plateau}}.
\newblock \bibinfo{journal}{Advances in Meteorology} \bibinfo{volume}{2017}.
\newblock \DOIprefix\doi{10.1155/2017/9823865}.
\bibitem[{Lihua et~al.(2007)Lihua, Yanben and Zhiqiang}]{Lihua2007}
\bibinfo{author}{Lihua, M.}, \bibinfo{author}{Yanben, H.}, \bibinfo{author}{Zhiqiang, Y.}, \bibinfo{year}{2007}.
\newblock \bibinfo{title}{{The possible influence of solar activity on Indian summer monsoon rainfall}}.
\newblock \bibinfo{journal}{Appl. Geophys.} \bibinfo{volume}{4}, \bibinfo{pages}{231--237}.
\newblock \DOIprefix\doi{10.1007/s11770-007-0029-4}.
\bibitem[{Malik and Br{\"{o}}nnimann(2018)}]{Malik2018}
\bibinfo{author}{Malik, A.}, \bibinfo{author}{Br{\"{o}}nnimann, S.}, \bibinfo{year}{2018}.
\newblock \bibinfo{title}{{Factors affecting the inter-annual to centennial timescale variability of Indian summer monsoon rainfall}}.
\newblock \bibinfo{journal}{Clim. Dyn.} \bibinfo{volume}{50}, \bibinfo{pages}{4347--4364}.
\newblock \DOIprefix\doi{10.1007/s00382-017-3879-3}.
\bibitem[{Mauas et~al.(2011)Mauas, Buccino and Flamenco}]{Mauas2011}
\bibinfo{author}{Mauas, P.J.}, \bibinfo{author}{Buccino, A.P.}, \bibinfo{author}{Flamenco, E.}, \bibinfo{year}{2011}.
\newblock \bibinfo{title}{{Long-term solar activity influences on South American rivers}}.
\newblock \bibinfo{journal}{J. Atmos. Sol.-Terr. Phys.} \bibinfo{volume}{73}, \bibinfo{pages}{377--382}.
\newblock \DOIprefix\doi{10.1016/j.jastp.2010.02.019}, \href{http://arxiv.org/abs/1003.0414}{{\tt arXiv:1003.0414}}.
\bibitem[{Mitchell et~al.(1979)Mitchell, Stockton and Meko}]{mitchell1979}
\bibinfo{author}{Mitchell, J.M.}, \bibinfo{author}{Stockton, C.W.}, \bibinfo{author}{Meko, D.M.}, \bibinfo{year}{1979}.
\newblock \bibinfo{title}{Evidence of a 22-year rhythm of drought in the western united states related to the hale solar cycle since the 17th century}, in: \bibinfo{editor}{McCormac, B.M.}, \bibinfo{editor}{Seliga, T.A.} (Eds.), \bibinfo{booktitle}{Solar-Terrestrial Influences on Weather and Climate}, \bibinfo{publisher}{Springer Netherlands}, \bibinfo{address}{Dordrecht}. pp. \bibinfo{pages}{125--143}.
\bibitem[{Nitka and Burnecki(2019)}]{Nitka2019}
\bibinfo{author}{Nitka, W.}, \bibinfo{author}{Burnecki, K.}, \bibinfo{year}{2019}.
\newblock \bibinfo{title}{{Impact of solar activity on precipitation in the United States}}.
\newblock \bibinfo{journal}{Physica A: Statistical Mechanics and its Applications} \bibinfo{volume}{527}, \bibinfo{pages}{121387}.
\bibitem[{Pai et~al.(2014)Pai, Sridhar, Rajeevan, Sreejith, Satbhai and Mukhopadhyay}]{Pai2014}
\bibinfo{author}{Pai, D.S.}, \bibinfo{author}{Sridhar, L.}, \bibinfo{author}{Rajeevan, M.}, \bibinfo{author}{Sreejith, O.P.}, \bibinfo{author}{Satbhai, N.S.}, \bibinfo{author}{Mukhopadhyay, B.}, \bibinfo{year}{2014}.
\newblock \bibinfo{title}{{Development of a new high spatial resolution (0.25° × 0.25°) long period (1901-2010) daily gridded rainfall data set over India and its comparison with existing data sets over the region}}.
\newblock \bibinfo{journal}{Mausam} \bibinfo{volume}{65}, \bibinfo{pages}{1--18}.
\newblock \DOIprefix\doi{10.54302/mausam.v65i1.851}.
\bibitem[{Rampelotto et~al.(2012)Rampelotto, Rigozo, da~Rosa, Prestes, Frigo, {Souza Echer} and Nordemann}]{Rampelotto2012}
\bibinfo{author}{Rampelotto, P.H.}, \bibinfo{author}{Rigozo, N.R.}, \bibinfo{author}{da~Rosa, M.B.}, \bibinfo{author}{Prestes, A.}, \bibinfo{author}{Frigo, E.}, \bibinfo{author}{{Souza Echer}, M.P.}, \bibinfo{author}{Nordemann, D.J.}, \bibinfo{year}{2012}.
\newblock \bibinfo{title}{{Variability of rainfall and temperature (1912-2008) parameters measured from Santa Maria (29°41'S, 53°48'W) and their connections with ENSO and solar activity}}.
\newblock \bibinfo{journal}{J. Atmos. Sol.-Terr. Phys.} \bibinfo{volume}{77}, \bibinfo{pages}{152--160}.
\newblock \DOIprefix\doi{10.1016/j.jastp.2011.12.012}.
\bibitem[{Rimbu et~al.(2021)Rimbu, Lohmann, Ionita, Czymzik and Brauer}]{Rimbu2021}
\bibinfo{author}{Rimbu, N.}, \bibinfo{author}{Lohmann, G.}, \bibinfo{author}{Ionita, M.}, \bibinfo{author}{Czymzik, M.}, \bibinfo{author}{Brauer, A.}, \bibinfo{year}{2021}.
\newblock \bibinfo{title}{{Interannual to millennial-scale variability of River Ammer floods and its relationship with solar forcing}}.
\newblock \bibinfo{journal}{Int J Climatol .} \bibinfo{volume}{41}, \bibinfo{pages}{E644--E655}.
\newblock \DOIprefix\doi{10.1002/joc.6715}.
\bibitem[{Selvaraj and Aditya(2011)}]{selvaraj2011study}
\bibinfo{author}{Selvaraj, R.S.}, \bibinfo{author}{Aditya, R.}, \bibinfo{year}{2011}.
\newblock \bibinfo{title}{Study on correlation between southwest and northeast monsoon rainfall over tamil nadu.}
\newblock \bibinfo{journal}{Univers. J. Environ. Res. Technol.} \bibinfo{volume}{1}.
\bibitem[{Selvaraj and Aditya(2012)}]{selvaraj2012}
\bibinfo{author}{Selvaraj, R.S.}, \bibinfo{author}{Aditya, R.}, \bibinfo{year}{2012}.
\newblock \bibinfo{title}{{The solar influence on the monsoon rainfall over Tamil Nadu}}.
\newblock \bibinfo{journal}{J. Ind. Geophys. Union} \bibinfo{volume}{16}, \bibinfo{pages}{107--111}.
\bibitem[{Selvaraj et~al.(2009a)Selvaraj, Muthuchami and Nancharaiah}]{Selvaraj2009}
\bibinfo{author}{Selvaraj, R.S.}, \bibinfo{author}{Muthuchami, A.}, \bibinfo{author}{Nancharaiah, M.}, \bibinfo{year}{2009}a.
\newblock \bibinfo{title}{{Influence of sunspot activity on the annual rainfall of Tamil Nadu, India}}.
\newblock \bibinfo{journal}{Indian J. Phys.} \bibinfo{volume}{83}, \bibinfo{pages}{1251--1258}.
\newblock \DOIprefix\doi{10.1007/s12648-009-0106-z}.
\bibitem[{Selvaraj et~al.(2009b)Selvaraj, Muthuchami and Nancharaiah}]{selvaraj2009influence}
\bibinfo{author}{Selvaraj, R.S.}, \bibinfo{author}{Muthuchami, A.}, \bibinfo{author}{Nancharaiah, M.}, \bibinfo{year}{2009}b.
\newblock \bibinfo{title}{{Influence of sunspot activity on the annual rainfall of Tamil Nadu, India}}.
\newblock \bibinfo{journal}{Indian J. Phys.} \bibinfo{volume}{83}, \bibinfo{pages}{1251--1258}.
\newblock \DOIprefix\doi{10.1007/s12648-009-0106-z}.
\bibitem[{Selvaraj et~al.(2013)Selvaraj, Umarani, Mahalakshmi and May}]{Selvaraj2013}
\bibinfo{author}{Selvaraj, R.S.}, \bibinfo{author}{Umarani, R.}, \bibinfo{author}{Mahalakshmi, N.}, \bibinfo{author}{May, M.A.}, \bibinfo{year}{2013}.
\newblock \bibinfo{title}{{Correlative study on Solar activity and all India rainfall : Cycle to Cycle Analysis}}.
\newblock \bibinfo{journal}{J. Ind. Geophys. Union} \bibinfo{volume}{17}, \bibinfo{pages}{59--63}.
\bibitem[{Song et~al.(2022)Song, Li, Gu and Xiao}]{yan2022}
\bibinfo{author}{Song, Y.}, \bibinfo{author}{Li, Z.}, \bibinfo{author}{Gu, Y.}, \bibinfo{author}{Xiao, Z.}, \bibinfo{year}{2022}.
\newblock \bibinfo{title}{Impact of solar activity on snow cover variation over the tibetan plateau and linkage to the summer precipitation in china}.
\newblock \bibinfo{journal}{Front. Earth Sci.} \bibinfo{volume}{9}.
\newblock \DOIprefix\doi{10.3389/feart.2021.756762}.
\bibitem[{Soon et~al.(1996)Soon, Posmentier and Baliunas}]{Soon1996}
\bibinfo{author}{Soon, W.H.}, \bibinfo{author}{Posmentier, E.S.}, \bibinfo{author}{Baliunas, S.L.}, \bibinfo{year}{1996}.
\newblock \bibinfo{title}{{Inference of Solar Irradiance Variability from Terrestrial Temperature Changes, 1880–1993: An Astrophysical Application of the Sun‐Climate Connection}}.
\newblock \bibinfo{journal}{Astrophys. J.} \bibinfo{volume}{472}, \bibinfo{pages}{891--902}.
\newblock \DOIprefix\doi{10.1086/178119}.
\bibitem[{Stager et~al.(2007)Stager, Ruzmaikin, Conway, Verburg and Mason}]{Stager2007}
\bibinfo{author}{Stager, J.C.}, \bibinfo{author}{Ruzmaikin, A.}, \bibinfo{author}{Conway, D.}, \bibinfo{author}{Verburg, P.}, \bibinfo{author}{Mason, P.J.}, \bibinfo{year}{2007}.
\newblock \bibinfo{title}{{Sunspots, El Ni{\~{n}}o, and the levels of Lake Victoria, East Africa}}.
\newblock \bibinfo{journal}{J. Geophys. Res. Atmos.} \bibinfo{volume}{112}, \bibinfo{pages}{1--13}.
\newblock \DOIprefix\doi{10.1029/2006JD008362}.
\bibitem[{Svensmark(2007)}]{Svensmark2007}
\bibinfo{author}{Svensmark, H.}, \bibinfo{year}{2007}.
\newblock \bibinfo{title}{Cosmoclimatology: a new theory emerges}.
\newblock \bibinfo{journal}{Astron. Geophys.} \bibinfo{volume}{48}, \bibinfo{pages}{18--24}.
\newblock \DOIprefix\doi{10.1111/j.1468-4004.2007.48118.x}.
\bibitem[{Tapping and Charrois(1994)}]{tapping1994limits}
\bibinfo{author}{Tapping, K.}, \bibinfo{author}{Charrois, D.}, \bibinfo{year}{1994}.
\newblock \bibinfo{title}{Limits to the accuracy of the 10.7 cm flux}.
\newblock \bibinfo{journal}{Sol. Phys.} \bibinfo{volume}{150}, \bibinfo{pages}{305--315}.
\bibitem[{Tapping(2013)}]{tapping201310}
\bibinfo{author}{Tapping, K.F.}, \bibinfo{year}{2013}.
\newblock \bibinfo{title}{{The 10.7 cm solar radio flux (F10.7)}}.
\newblock \bibinfo{journal}{Space Weather} \bibinfo{volume}{11}, \bibinfo{pages}{394--406}.
\newblock \DOIprefix\doi{10.1002/swe.20064}.
\bibitem[{Thomas and Abraham(2022a)}]{thomas2022impact}
\bibinfo{author}{Thomas, E.}, \bibinfo{author}{Abraham, N.P.}, \bibinfo{year}{2022}a.
\newblock \bibinfo{title}{Impact of solar activity on the seasonal rainfall of kerala, india}.
\newblock \bibinfo{journal}{India (January 26, 2022)} \DOIprefix\doi{dx.doi.org/10.2139/ssrn.4102224}.
\bibitem[{Thomas and Abraham(2022b)}]{Thomas2022}
\bibinfo{author}{Thomas, E.}, \bibinfo{author}{Abraham, N.P.}, \bibinfo{year}{2022}b.
\newblock \bibinfo{title}{{Relationship between sunspot number and seasonal rainfall over Kerala using wavelet analysis}}.
\newblock \bibinfo{journal}{J. Atmos. Sol.-Terr. Phys} \bibinfo{volume}{240}, \bibinfo{pages}{105943}.
\newblock \DOIprefix\doi{10.1016/j.jastp.2022.105943}.
\bibitem[{Thomas et~al.(2023)Thomas, Joseph and Abraham}]{Thomas2023}
\bibinfo{author}{Thomas, E.}, \bibinfo{author}{Joseph, I.}, \bibinfo{author}{Abraham, N.P.}, \bibinfo{year}{2023}.
\newblock \bibinfo{title}{{Wavelet analysis of annual rainfall over Kerala and sunspot number}}.
\newblock \bibinfo{journal}{New Astron} \bibinfo{volume}{98}.
\newblock \DOIprefix\doi{10.1016/j.newast.2022.101944}.
\bibitem[{Tiwari and Kumar(2018)}]{tiwari2018}
\bibinfo{author}{Tiwari, B.}, \bibinfo{author}{Kumar, M.}, \bibinfo{year}{2018}.
\newblock \bibinfo{title}{The solar flux and sunspot number; a long-trend analysis}.
\newblock \bibinfo{journal}{International Annals of Science} \bibinfo{volume}{5}, \bibinfo{pages}{47--51}.
\newblock \DOIprefix\doi{10.21467/ias.5.1.47-51}.
\bibitem[{Tiwari et~al.(2021)Tiwari, Xu, Adhikari and Chapagain}]{tiwari2021}
\bibinfo{author}{Tiwari, B.}, \bibinfo{author}{Xu, J.}, \bibinfo{author}{Adhikari, B.}, \bibinfo{author}{Chapagain, N.}, \bibinfo{year}{2021}.
\newblock \bibinfo{title}{Wavelet and cross correlation analysis on some climatology parameters of nepal}.
\newblock \bibinfo{journal}{BIBECHANA} \bibinfo{volume}{18}, \bibinfo{pages}{105--116}.
\newblock \DOIprefix\doi{10.3126/bibechana.v18i2.33805}.
\bibitem[{Tsiropoula(2003)}]{Tsiropoula2003}
\bibinfo{author}{Tsiropoula, G.}, \bibinfo{year}{2003}.
\newblock \bibinfo{title}{{Signatures of solar activity variability in meteorological parameters}}.
\newblock \bibinfo{journal}{J. Atmos. Sol.-Terr. Phys.} \bibinfo{volume}{65}, \bibinfo{pages}{469--482}.
\newblock \DOIprefix\doi{10.1016/S1364-6826(02)00295-X}.
\bibitem[{Usoskin(2017)}]{Usoskin2017}
\bibinfo{author}{Usoskin, I.G.}, \bibinfo{year}{2017}.
\newblock \bibinfo{title}{{A History of Solar Activity over Millennia}}.
\newblock \bibinfo{journal}{Living Rev. Solar Phys} \bibinfo{volume}{14}, \bibinfo{pages}{3}.
\newblock \DOIprefix\doi{https://doi.org/10.1007/s41116-017-0006-9}.
\bibitem[{Vaquero(2004)}]{Vaquero2004SolarSI}
\bibinfo{author}{Vaquero, J.M.}, \bibinfo{year}{2004}.
\newblock \bibinfo{title}{Solar signal in the number of floods recorded for the tagus river basin over the last millennium}.
\newblock \bibinfo{journal}{Climatic Change} \bibinfo{volume}{66}, \bibinfo{pages}{23--26}.
\bibitem[{Warrier et~al.(2017)Warrier, Sandeep and Shankar}]{Warrier2017}
\bibinfo{author}{Warrier, A.K.}, \bibinfo{author}{Sandeep, K.}, \bibinfo{author}{Shankar, R.}, \bibinfo{year}{2017}.
\newblock \bibinfo{title}{{Climatic periodicities recorded in lake sediment magnetic susceptibility data: Further evidence for solar forcing on Indian summer monsoon}}.
\newblock \bibinfo{journal}{Geosci. Front.} \bibinfo{volume}{8}, \bibinfo{pages}{1349--1355}.
\newblock \DOIprefix\doi{10.1016/j.gsf.2017.01.004}.
\bibitem[{Wasko and Sharma(2009)}]{Wasko2009}
\bibinfo{author}{Wasko, C.}, \bibinfo{author}{Sharma, A.}, \bibinfo{year}{2009}.
\newblock \bibinfo{title}{{Effect of solar variability on atmospheric moisture storage}}.
\newblock \bibinfo{journal}{Geophys. Res. Lett.} \bibinfo{volume}{36}.
\newblock \DOIprefix\doi{10.1029/2008GL036310}.
\bibitem[{Wirth et~al.(2013)Wirth, Glur, Gilli and Anselmetti}]{Wirth2013}
\bibinfo{author}{Wirth, S.B.}, \bibinfo{author}{Glur, L.}, \bibinfo{author}{Gilli, A.}, \bibinfo{author}{Anselmetti, F.S.}, \bibinfo{year}{2013}.
\newblock \bibinfo{title}{{Holocene flood frequency across the Central Alps - solar forcing and evidence for variations in North Atlantic atmospheric circulation}}.
\newblock \bibinfo{journal}{Quat. Sci. Rev.} \bibinfo{volume}{80}, \bibinfo{pages}{112--128}.
\newblock \URLprefix \url{http://dx.doi.org/10.1016/j.quascirev.2013.09.002}, \DOIprefix\doi{10.1016/j.quascirev.2013.09.002}.
\bibitem[{Yu et~al.(2019)Yu, Wang, Yu and Kang}]{YU2019}
\bibinfo{author}{Yu, X.}, \bibinfo{author}{Wang, Y.}, \bibinfo{author}{Yu, S.}, \bibinfo{author}{Kang, Z.}, \bibinfo{year}{2019}.
\newblock \bibinfo{title}{Synchronous droughts and floods in the southern chinese loess plateau since 1646 ce in phase with decadal solar activities}.
\newblock \bibinfo{journal}{Glob Planet Change} \bibinfo{volume}{183}, \bibinfo{pages}{103033}.
\newblock \DOIprefix\doi{https://doi.org/10.1016/j.gloplacha.2019.103033}.
\bibitem[{Zhai(2017)}]{Zhai2017}
\bibinfo{author}{Zhai, Q.}, \bibinfo{year}{2017}.
\newblock \bibinfo{title}{{Influence of solar activity on the precipitation in the North-central China}}.
\newblock \bibinfo{journal}{New Astron.} \bibinfo{volume}{51}, \bibinfo{pages}{1339--1351}.
\newblock \DOIprefix\doi{10.1016/j.newast.2016.09.003}.
\bibitem[{Zhao et~al.(2004)Zhao, Han and Li}]{ZhaoJuanHanYan-BenandLi2004}
\bibinfo{author}{Zhao, J.}, \bibinfo{author}{Han, Y.B.}, \bibinfo{author}{Li, Z.A.}, \bibinfo{year}{2004}.
\newblock \bibinfo{title}{{The Effect of Solar Activity on the Annual Precipitation in the Beijing Area}}.
\newblock \bibinfo{journal}{Chin. J. Astron. and Astrophys.} \bibinfo{volume}{4}, \bibinfo{pages}{189--197}.

\end{thebibliography}






\end{document}